\documentclass[pre,preprint,showpacs,preprintnumbers,amsmath,amssymb]{revtex4}

\usepackage{graphicx}

\usepackage{dcolumn}
\usepackage{bm}
\usepackage[mathscr]{eucal}
\usepackage{mathrsfs}

\newtheorem{Theorem}{Theorem}
\newtheorem{Definition}{Definition}

\begin{document}
\preprint{ver. 2.0}

\title{A thermodynamic description of the glass state and its application to glass transition}

\author{Koun Shirai}
\affiliation{%
The Institute of Scientific and Industrial Research, Osaka University, 8-1 Mihogaoka, Ibaraki, Osaka 567-0047, Japan
}%

\begin{abstract}
Many properties of solids, such as the glass state, hysteresis, and memory effects, are commonly treated as nonequilibrium phenomena, which involve many conceptual difficulties. 
However, few studies have addressed the problem of understanding equilibrium itself. 
Equilibrium is commonly assessed based on the assumption that its thermodynamic state should be determined solely by temperature and pressure. However, this assumption must be fundamentally reappraised from the beginning through a rigorous definition of equilibrium because no rigorous proof for this assumption exists.
Previous work showed that for solids, the equilibrium positions of all constituent atoms of the solid are state variables ({\it i.e.}, thermodynamic coordinates, or ``TCs"). 
In this study, this conclusion is further elaborated starting from the principles of solid-state physics. 
The internal variables such as the fictive temperature qualify as TCs on this ground, if suitably treated.
This theory is applied to glass materials, for which many challenges remain.
Results show that first, the glass state is an equilibrium state. Accordingly, the properties of a glass can be solely described by the present positions of the atom, irrespective of their previous history, which is consistent with the definition of a state in the thermodynamic context.
Second, the glass transition, although a nonequilibrium phenomenon, is well described using TCs if the thermal part of the energy is be assumed to be well separated from the structural part. The only dynamic parameter involved in this approach is the relaxation time, which is uniquely determined using the present values of the TCs. Therefore, complicated functions describing the past history, which are widely used in the glass literature, are unnecessary. 
This implies that the activation energy for the structural relaxation strongly depends on TCs. 
This finding provides a reasonable understanding of the large deviations from the Arrhenius law, which often occur in glasses. The unrealistic values of the activation energy and of the pre-exponential factor of the relaxation time can be resolved on this basis.
The theory is particularly suitable for experiments that do not involve hypothetical quantities such as the effective temperature or hypothetical models such as the ideal glass model; therefore, all quantities are measurable.

\end{abstract}

\pacs{1.0}

\maketitle


\section{Introduction}
\label{sec:intro}
\paragraph{Background.} 
Thermodynamics is an elemental discipline underpinning the developments of a wide range of scientific and engineering fields. The principles of thermodynamics are robust and are not invalidated by the advent of quantum mechanics. Despite this universal characteristic, severe restrictions are imposed on applying thermodynamics to investigate the properties of solids. Many problems arise from the conceptual difficulty of defining state variables.
This is easy to answer in an introductory course on thermodynamics, which teaches us that for gases, equilibrium states are described solely by two state variables: temperature $T$ and pressure $p$. Many researchers believe that the same holds true for solids. 
(When chemical reactions are considered, the number of the chemical species and the numbers of their atoms appear as state variables. Here, the issue is whether further state variables exist even subsequent to fixing these numbers. Furthermore, when electromagnetic properties are considered, electric polarization and magnetization appear as state variables. These two variables are not essential for the present study, and therefore here electromagnetic properties are disregarded.)
As later discussed, many phenomena in solids cannot be described solely using the two variables $T$ and $p$. For example, the physical properties of plastics cannot be described solely using $T$ and $p$, and the mechanical properties of metals vary depending on heat treatments. These difficulties have been averted by considering these phenomena to be nonequilibrium and no longer treating them within the framework of thermodynamics. However, all solids exhibit hysteresis to a certain degree. If thermodynamics were unable to explain such commonly observed phenomena, it would risk being relegated from the top ranks of physics principles. 

Bridgman was the first to confront the hysteresis issue from the thermodynamics viewpoint when he stated:
``But the admission of general impotence in the presence of irreversible processes appears on reflection to be a surprising thing. Physics does not usually adopt such an attitude of defeatism" (\cite{Bridgman41}, p.~133). The present author advocates Bridgman's viewpoint.
His approach to describing plastic deformation is remarkable \cite{Bridgman50}; he treats stress and strain as independent variables---in elastic theory these are interrelated with each other. Unfortunately, Bridgman's theory has not been further developed.

No rigorous proof is available that $T$ and $p$ are the only independent state variables for solids. However, despite this lack of proof, this proposition is decisive when deciding whether a property is an equilibrium property. In view of the importance of this proposition in a wide range of applications, it must be critically examined. To this end, we must first define equilibrium; however, this requires understanding what state variables are. Thus, the argument becomes circular, which makes a coherent definition of equilibrium difficult to obtain \cite{comment-equilibrium}. 
This logical difficulty was resolved by Gyftopoulos and Beretta, who provided a coherent framework of thermodynamics \cite{Gyftopoulos}.
The basis of their theory relies on the fact that numerous equilibrium states exist for a given system, whereas one and only one stable equilibrium state exists for a given set of constraints. The latter part of the preceding sentence is an alternative expression of the second law of thermodynamics. Every equilibrium state is classified by constraints. It is observed that a full set of constraints is required to justify the second law. 

Previous work examined the consequences of applying this theory of equilibrium to solids \cite{StateVariable}. Hereinafter, the term ``thermodynamic coordinates" (TCs) is used for state variables, following Zemanski \cite{Zemansky}. The result is that the time-averaged positions of all atoms comprising a solid are necessary and sufficient TCs for the solid. 
At first glance, this conclusion seems strange, because too many TCs are needed for a thermodynamic description, which seems to be in conflict with the spirit of thermodynamics {(\it i.e.}, describing many-particle systems with a minimum of variables) \cite{Callen}. However, the most important characteristics of a TC are {\em definiteness} and {\em distinguishability}, which are explained below.

As opposed to a gas, a solid can exist in a variety of equilibrium states. For example, displacement of a single atom in a solid to an interstitial site creates another state of the solid, which is an equilibrium state because the defect structure does not vary unless it is annealed out. The thermodynamic properties of the state thus created differ from those of a perfect crystal because the displacement modifies the phonon spectrum, which in turn alters the specific heat of the crystal. In fact, an overwhelming number of atom displacements are possible, each of which corresponds to a distinct equilibrium state.
The definiteness mentioned above implies that a unique value of TC can be attached to each equilibrium state. For a gas in equilibrium, there is no unique position for each atom. Only volume $V$ has a unique value, provided $T$ is fixed. Conversely, the atoms in a solid have definite average positions, that is, equilibrium positions, which implies that the time-averaged positions of atoms describe the equilibrium states of the solid. 
The distinguishability means that every atom can be distinguished by its equilibrium position from all other atoms. For a gas, all atoms share the same volume and cannot be distinguished by the time-averaged positions. 

The fact that a large number of atom positions exist does not conflict with the principles of thermodynamics. In the first law of thermodynamics,
\begin{equation}
dU = TdS - \sum_{j=1}^{M} F_{j} \cdot X_{j},
\label{eq:first-law}
\end{equation}
nothing restricts the number $M$ of state variables $\{ X_{j} \}$. Here, $U$ is the internal energy of the system, $S$ is the entropy of the system, and $F_{j}$ is the generalized force corresponding to $X_{j}$. 
The microscopic nature of atom positions is irrelevant for being a TC; the word ``microscopic" only makes sense when viewed from the human scale. The universal laws of physics must hold independent of our scale. The above definitions of equilibrium and TCs are not a matter of interpretation. They are a logical and rigorous consequence of the second law of thermodynamics. 

\paragraph{Purpose of this paper.} 
This study applies this different view of equilibrium to the glass problem, which is one of the most interesting applications of the theory. 
In the early days, glass confronted physicists with a serious difficulty because it seemed to contradict the third law of thermodynamics. Glasses have nonzero entropy even when approaching $T=0$. Simon and contemporaries explained this contradiction by deeming the glass state to be a nonequilibrium state \cite{Simon30, Fowler-Guggenheim}. Since then, numerous studies over the past century used this explanation \cite{Davies53a, Jackle86, Angell99, Rao02,Mysen-Richet05,Berthier11,Biroli13,Berthier16}.
However, conceptual problems remain unsolved: the nature of the order parameter in glasses, the Kauzmann paradox \cite{Kauzmann48}, ideal glasses, and the Prigogin-Defay relation are still being debated. All these issues are deeply related to thermodynamic equilibrium.

To address these issues, the present work describes the glass state from a different view of thermodynamics. Briefly speaking, all static states are equilibrium states, leading to the conclusion that glasses are an equilibrium state once solidification is completed, which significantly differs from the widely held view that glasses are in a nonequilibrium state. Next, a thermodynamic description of the glass transition is provided, focusing on the specific heat $C$ versus $T$ (throughout this paper, $C$ is taken to mean the isobaric specific heat, so that the subscript $p$ is omitted).
Glass transitions are nonequilibrium phenomena, because, by definition, transitions are time dependent. Despite this, thermodynamics proves quite useful for analyzing processes. A significant advantage of the present theory is that, under appropriate conditions (which we later call the adiabatic approximation of the second kind), the transitions can be described based only on the present state, so complicated functions of past history are not required. An important outcome of this approach is that the activation energy is a strong function of enthalpy and therefore of temperature. This provides useful insights for studying many solid-state phenomena that are presently considered as nonequilibrium phenomena, such as the memory function of phase-change materials \cite{MRS-PCM19} and the aging and rejuvenating effect of spin glasses \cite{note1}. 

This paper is organized as follows: 
Section II summarizes the general framework of thermodynamics based on this different conception of equilibrium, which connects the work of Ref.~\cite{StateVariable} with the present discussion. In addition, Sec. II provides a microscopic derivation of the TCs of solids and the hierarchy of approximations for thermodynamic descriptions.
Section III describes the glass state. A major difference with the conventional view held by most who study glasses is that, once solidified, glasses are indeed in equilibrium. 
Section~IV presents a concrete implementation of the theory for analyzing a $C$-$T$ curve.
Section~V gives a useful interpretation of the activation energy, which resolves some of the current problems of glasses, and Sec. VI concludes the study.

\section{Grand work on the thermodynamics of solids}
\label{sec:TDfundaments}

\subsection{Thermodynamic equilibrium}
\label{sec:equilibrium}
\setcounter{paragraph}{0}
\paragraph{Equilibrium and thermodynamic coordinates.}
We begin by defining equilibrium. 
Although no essential differences should exist in the thermodynamics principles between gases and solids, an unambiguous definition of equilibrium is required for solids because of the variety of equilibria possible in solids and the many difficulties that arise therefrom.
The existence of numerous equilibria combined with the second law of thermodynamics clouds the picture of stable equilibrium, making it difficult to understand (Ref.~\cite{Gyftopoulos}, p.~63). 
For a gas, the shape of the container in which the gas is enclosed is irrelevant to the thermodynamic properties of the gas, provided the volume is the same. Therefore, the shape is not a TC. 
In contrast, a solid can be deformed in various ways, in addition to homogeneous deformations, and all deformations alter the internal energy of the solid. Hence, the work part of Eq.~(\ref{eq:first-law}) is expressed by $-V \int \sigma({\mathbf r}) d\varepsilon({\mathbf r})$, where $\sigma$ and $\varepsilon$ are the stress and strain, respectively. The strain at each point in the solid is regarded as a TC. This implies that the number of the TCs, $M$ in Eq.~(\ref{eq:first-law}), is virtually infinite.

It is impossible to describe equilibrium without specifying constraints. Gibbs noticed this role of constraints---passive resistance in his words---more than one century ago \cite{Gibbs}, and equilibrium holds only within the given constraints. However, at the time, the nature of constraints was unknown. Today, we can answer this question: constraints are no more than energy barriers of any kind. 
Another problem that has not been sufficiently addressed, is the relationship between constraints and TCs. Reiss may have been the first to realize that a one-to-one correspondence exists between a constraint $\xi_{j}$ and a TC $X_{j}$ \cite{Reiss}. A constraint specifies the range of a quantity $x$ ({\it i.e.}, the position, energy, chemical species, or any other observable) that the particles can visit. At a finite $T$, the quantity $x(t)$ accompanies fluctuations within the constraint as a function of time $t$. A time-averaged value of $x(t)$ over a period $t_{0}$,
\begin{equation}
X_{j} = \frac{1}{t_{0}} \int_{0}^{t_{0}} x_{j}(t) dt,
\label{eq:time-average}
\end{equation}
is fixed in an equilibrium state. In this manner, a constraint creates a TC.

The remaining problem is how to define equilibrium without using TC as a predefined quantity. 
This problem was resolved by Gyftopoulos and Beretta \cite{Gyftopoulos}, who defined equilibrium as follows:
\begin{Definition}[Thermodynamic equilibrium] \label{def:equilibrium}
It is impossible to change the stable equilibrium state of a system to any other state with the sole effect on the environment being a raise of the weight.
\end{Definition}
In short, the raise of the weight means performing positive work on the environment. This manner of defining equilibrium does not require knowledge of the state of the system in question. We can treat the interior of the system as a black box; knowledge of how the system affects the environment is the sole requirement. This definition of equilibrium is consistent with an intuitive understanding of equilibrium: only static states are equilibrium states.  
Numerous equilibrium states exist for a given system. By combing Def.~\ref{def:equilibrium} to the expression of the second law that {\it one and only one stable equilibrium state exists for a given set of constraints}, we find that a system is characterized by a full set of constraints. Constraints include everything to determine the structures of a material.

If one constraint is altered, the equilibrium state changes. The properties of a solid change when the atom configuration changes. Displacing one atom changes the properties of the solid, which is exploited for memory and switching devices in electronics. A phase transition that involves displacement produces different properties of a solid; for example, ferroelectric materials have different polarizations depending on the applied electric field. A collective displacement of atoms creates dislocations, which produce different mechanical properties. These defect states are in fact equilibrium states, provided they remain static. If a defect state were not an equilibrium state, the solid could perform work on the environment without altering the constraints. Performing work is possible only by compensating it with a decrease in the internal energy of the solid, which is tantamount to obtaining work by cooling a system. This contradicts the second law, which prohibits perpetual-motion machines of the second kind. Therefore, defect states must be equilibrium states.

Upon analyzing the examples above, we find that the positions of all the constituent atoms, $\{ {\mathbf R}_{j} \}_{j=1, \dots, N_{\rm at}}$, where $N_{\rm at}$ is the number of atoms, must be known for solids. At finite temperature $T$, the position of atoms in a solid depends strongly on time $t$. An instantaneous position ${\mathbf R}_{j}(t)$ cannot be a TC because it does not give a unique value for each equilibrium state. Only the time-averaged value $\bar{\mathbf R}_{j}$ is unique. Therefore, the set of time-averaged atom positions $\{ \bar{\mathbf R}_{j} \}_{j=1, \dots, N_{\rm at}}$ ({\it i.e.}, equilibrium positions) are the TCs for a solid. Moreover, these variables form a complete set of TCs that fully describe the thermodynamic properties of the solid.
A more detailed discussion of this subject is available in Ref.~\cite{StateVariable}.

The above argument does not depend on the type of structure of the solid. Whether it has a periodicity of a lattice is irrelevant. If we take snapshots of a gas and a glass, the atomic arrangements within the gas and the glass have the same sense of ``randomness" in each snapshot. However, for a gas, the instantaneous positions of atoms are irrelevant to the thermodynamic properties of the gas. For a gas, time-averaging the atom positions in Eq.~(\ref{eq:time-average}) destroys entirely the distinguishability of atom positions, leaving only the volume as a relevant quantity for thermodynamic properties. 
We can say that missing information occurs in terms of information theory \cite{Ben-Naim,Haar-Thermostat,Rosenkrantz83}. Conversely, for the glass case, time-averaging does not destroy the distinguishability of atom positions. The correlation between different atom positions $\bar{\mathbf R}_{j}$ and $\bar{\mathbf R}_{i}$ does not change with time.
This fact explains why it makes sense to describe the thermodynamic properties of a glass by using the time-averaged atom positions $\{ \bar{\mathbf R}_{j} \}$ as the TCs.

\paragraph{Timescale issues.} 
Let us identify two types of equilibria, which are needed in what follows.
System $A$ interacts with the environment in two ways: thermal interactions by exchanging heat $Q$ and mechanical interaction by exchanging work $W$, as described by Eq.~(\ref{eq:first-law}).
When an interaction with the environment is turned on, the state of the system changes to a new state. When the state of the system ceases to change, the system is in equilibrium. 
When zero net heat is transferred to the environment, the equilibrium is referred to as {\it thermal equilibrium}. When zero net work is produced between the system and the environment, the equilibrium is referred to as {\it mechanical equilibrium}. If both interactions vanish, the equilibrium is referred to as {\it thermodynamic equilibrium}. The time required to reach thermal equilibrium is called the thermal relaxation time $\tau_{t}$, and the time required to reach mechanical equilibrium is called the mechanical relaxation time $\tau_{m}$. Usually, thermal and mechanical equilibria are established simultaneously, $\tau_{t} \approx \tau_{m}$, so that the distinction is unnecessary. However, in this study, the distinction becomes important. 

Constraint is a theoretical device to idealize energy barriers. It perfectly inhibits the change in a ``go or no go" manner (see page 10 in Ref.~\cite{Hatsopoulos}). However, because all energy barriers in real solids are finite, we can only speak of equilibrium within a given relaxation time.
In many applications, the distinction between stable (global) and metastable (local) equilibrium states is useful. 
However, in the present context, this distinction is not important. 
For example, a gas mixture of nitrogen and hydrogen is stable at ambient temperature but, in astronomical time, it will become ammonia gas. If nuclear reactions are taken into account, then nothing is stable except protons and electrons. 
Each constraint $\xi_{j}$ is thus associated with an energy barrier, which determines the corresponding relaxation time $\tau_{j}$. The mechanical relaxation time $\tau_{m}$ is the collective name for a set of $\{ \tau_{j} \}$.

\subsection{Hierarchy of thermodynamics in solids} 

\subsubsection{From solid-state theory to thermodynamics} 
\label{sec:electron-theory}

Thermodynamics itself does not provide recipes for deriving the formula for the internal energy $U$. This task falls to microscopic theories.
Today, the most reliable method to calculate the energy of solids at $T=0$ is density-functional theory (DFT) \cite{Callaway84,Parr-Yang89,Zangwill15}. 
The ground state of a given solid is uniquely determined by the electron density $\rho({\mathbf r})$ for a given external potential. For numerous problems of solids, the adiabatic approximation holds \cite{BornHuang}. Electronic and atomic coordinates are decoupled, because of the large difference in their masses. Based on this approximation, DFT restates that the ground-state energy $E_{\rm g.s.}$ of a solid is uniquely determined by the positions $\{ {\mathbf R}_{j} \}_{j=1, \dots, N_{\rm at}}$ of the atoms that comprise the solid. The ground-state energy $E_{\rm g.s.}$ can be expressed as a functional $E$ of the atom positions,
\begin{equation}
E_{\rm g.s.} = E[ \{ {\mathbf R}_{j} \} ].
\label{eq:dft}
\end{equation}
Throughout this paper, the spin freedom is not taken into consideration, as stated in Introduction. This relationship is valid even when atoms are not in their equilibrium positions; otherwise, virtually all {\it ab initio} molecular-dynamics simulations would lose their rigorous grounding.
{\it The ground-state energy $E_{\rm g.s.}(t)$ of a solid at the present time is uniquely determined solely by the present positions $\{ {\mathbf R}_{j}(t) \}$ of atoms, irrespective of their past history.} 

At finite temperature, atoms in a solid undergo rapid motion about their equilibrium positions, ${\mathbf R}_{j}(t) = \bar{\mathbf R}_{j} + {\mathbf u}_{j}(t)$, where $ {\mathbf u}_{j}(t)$ is a small displacement from the average position $\bar{\mathbf R}_{j}$. The characteristic time of atomic motion in solids is the phonon period $\tau_{\rm ph}$, which is of the order of 10 fs. Although this timescale is very short, it is longer than the response time of electrons, so that the thermal properties of solids can be described by DFT, despite it being a zero-temperature theory. By averaging over a timescale much longer than $\tau_{\rm ph}$, the averaged energy, that is, internal energy $U$, is given by
\begin{equation}
U = \bar{E}_{\rm g.s.}( \{ {\mathbf R}_{j}(t) \} ) = E_{\rm g.s.}( \{ \bar{\mathbf R}_{j} \} ) 
 + \sum_{k} \frac{1}{2} \omega_{k}^{2} \bar{q}_{k}^{2},
\label{eq:internalU-aveE}
\end{equation}
where $\bar{q}_{k}$ is the average of the amplitude of the $k$th normal mode of frequency $\omega_{k}$. Normal modes $\{ q_{k} \}$ are obtained by a unitary transformation of displacements $\{ u_{j} \}$, so $U$ is expressed as
\begin{equation}
U = U_{s}(\{ \bar{\mathbf R}_{j} \} ) + U_{t}(\{ \bar{\mathbf u}_{j} \} ).
\label{eq:UeqUpPlusPk}
\end{equation}
The first term $U_{s}$ on the right-hand side is the potential part of $U$, which corresponds to the first term in Eq.~(\ref{eq:internalU-aveE}) and is called the {\em structural part}. The second term $U_{t}$ is the kinetic part of $U$, which corresponds to the second term in Eq.~(\ref{eq:internalU-aveE}) and is called the {\em phonon part}.

The temperature dependence of $U$ stems from the phonon amplitude in the form $\frac{1}{2} \omega_{k}^{2} \bar{q}_{k}^{2} = (1/2+\bar{n}_{k} ) \hbar \omega_{k}$, where $\bar{n}_{k}$ is the Bose occupation number, which is given by $\bar{n}_{k} = (e^{\hbar \omega_{k} /k_{\rm B}T}-1)^{-1}$ ($\hbar$ is Planck's constant and $k_{\rm B}$ is Boltzmann's constant). The phonon frequency $\omega_{k}$ is determined by force constants, which in turn are determined by the curvature of $E_{\rm g.s.}( \{ \bar{\mathbf R}_{j} \} )$ with respect to $\bar{\mathbf R}_{j}$. 
Accordingly, $U_{t}$ is a function of $T$ and $\{ \mathbf{R}_{j} \}$, that is, $U_{t}(\{ \bar{\mathbf u}_{j} \} ) = U_{t}(T, \{ \bar{\mathbf R}_{j} \} )$.
Because the force constants usually depend only weakly on $\{ \mathbf{R}_{j} \}$, we can ignore this dependence. Consequently, the variables may be separated as follows:
\begin{equation}
U =U(T, \{ \bar{\mathbf R}_{j} \} ) = U_{t}(T) + U_{s}( \{ \bar{\mathbf R}_{j} \} ).
\label{eq:fundamental-relation1}
\end{equation}
When we measure the specific heat of solids, we normally obtain only the phonon part $U_{t}$, aside from the small contribution from the volume expansion. 
During phase transitions of crystals, $U_{s}$ appears as latent heat, so the structural part $U_{s}$ is not detected in specific-heat measurements. 
Usually, phase transitions occur on a timescale of the order of 1 ns. This structural relaxation time $\tau_{s}$ is longer than $\tau_{\rm ph}$, but remains very short compared with experimental timescales. Thus, in the usual crystallization-melting phase transition, we observe that thermal and mechanical equilibria are established on the same timescale.

\subsubsection{Fundamental relation of equilibrium}
\label{sec:connect-thermo}
To obtain the fundamental relation of equilibrium (FRE), we must know the entropy $S$.
It is elemental to calculate $S$ for a solid using the harmonic approximation:
\begin{equation}
S = S(\{ \bar{\mathbf R}_{j} \} ) =
 \sum_{k} \left\{ (\bar{n}_{k}+1)\ln(\bar{n}_{k}+1) -\bar{n}_{k} \ln \bar{n}_{k} \right\}.
\label{eq:SofPhonons}
\end{equation}
By inverting $T$ in Eq.~(\ref{eq:fundamental-relation1}) to $S$, we establish the FRE for a solid,
\begin{equation}
U =U(S, \{ \bar{\mathbf R}_{j} \} ).
\label{eq:fundamental-relation}
\end{equation}
If desired, the volume $V$ can be extracted from a set $\{ \bar{\mathbf R}_{j} \}$, and $V$ may be explicitly included as a TC. However, this variable is not of interest here and so is omitted from the following argumentation.
We have thus established the following theorem:
\begin{Theorem}[Properties of thermodynamic equilibrium] \label{th:first}
The thermodynamic properties of a solid at the present time are uniquely determined solely by the time-averaged present positions $\{ \bar{\mathbf R}_{j} \}$, irrespective of their past history. 
\end{Theorem}
The adjective ``present" preceding a word should be stressed. If this adjective were absent, the state of the solid would lose its meaning as a state in thermodynamics; namely, that a state is independent of the process in which it was obtained.
It is nonsensical to claim that the properties of a glass state depend on the process. Properties are states in the thermodynamic context \cite{Callen}. Obtaining different properties is achieved only because samples with different structures were obtained depending on the conditions of preparation. If two samples possessing the same properties are obtained, these two samples must be identified as the same state even if they obtained using different processes. In this manner, the glass state becomes a well-defined thermodynamic state.

Unfortunately, a hesitation exists in the community for using atom position as TC, because
researchers consider that the calculation of concrete forms of Eq.~(\ref{eq:SofPhonons}) and (\ref{eq:fundamental-relation}) belongs to the domain of statistical mechanical methods. However, this does not mean that using atom positions as TCs is a taboo in thermodynamics. 
The outcomes of statistical mechanics and thermodynamics must match each other: otherwise, one of the theories (or both) must be wrong.
In thermodynamics, $U(T)$ is obtained from experimental data on specific heat $C(T)$ as a function of $T$. 
In statistical mechanics, $U(T)$ is calculated from a microscopic quantity of the energy spectrum $g(u)$, provided that $g(u)$ is known by, for example, quantum theory. Between $U(T)$ and $g(u)$, there is a transformation, namely, a Laplace transformation. Hence, both ways are completely symmetric. The two theories are equivalent \cite{Mandelbrot64,Tisza63}.

\paragraph{Alternative expression of the fundamental relation of equilibrium.} 
All the properties of a solid in equilibrium are determined by the FRE (\ref{eq:fundamental-relation}). Any property $Y$ is given by a function of TCs as $Y=Y(S, \{ \bar{\mathbf R}_{j} \})$. Because $Y$ is a function of many variables, it is generally not invertible. Nonetheless, properties can be used as TCs, instead of $\{ \bar{\mathbf R}_{j} \}$. For example, internal energy $U$ is a function of $\{ \bar{\mathbf R}_{j} \}$. The relation between them is one to many. However, $\{ \bar{\mathbf R}_{j} \}$ can be grouped together by $U$. Let us denote $\{ \bar{\mathbf R}_{j} \}_{U}$ as a set of atom configurations that yield the same $U$. In this manner, a one-to-one correspondence between $U$ and $\{ \bar{\mathbf R}_{j} \}_{U}$ can be found. Within this grouping, $U$ can be used as a TC,
\begin{equation}
\left(T, \{ \bar{\mathbf R}_{j} \} \right) \rightarrow \left( T, U \right).
\label{eq:transTC1}
\end{equation}
Now, $T$ and $U$ are two independent variables, whereas for gas states one of the two is a dependent variable. The use by Bridgman of both stress and strain as independent variables \cite{Bridgman50} is supported on these grounds.
If each set $\{ \bar{\mathbf R}_{j} \}_{U}$ can be further grouped into subsets $\{ \bar{\mathbf R}_{j} \}_{U, Y_{2}}$ by another property $Y_{2}$, a set $\left( T, U, Y_{2} \right)$ forms a three-dimensional state space. In this manner, we can construct an unlimited-dimensional state space spanned by $\{T, U, Y_{2}, Y_{3}, \dots \}$.


\subsubsection{Adiabatic approximation of the second kind} 
\label{sec:second-adiabatic}
For glasses, an interesting situation occurs; namely, a large separation occurs between two relaxation times of $\tau_{t}$ (${\approx} \tau_{\rm ph}$) and $\tau_{s}$. The structural change in the glass transition occurs over a time longer than 1 s. This significant difference between $\tau_{t}$ and $\tau_{s}$ simplifies the treatment of thermal and mechanical equilibria, contrary to the common belief that treating the glass transition is complicated because of its nonequilibrium nature. Two types of equilibria can be treated separately. We call this separation {\em the adiabatic approximation of the second kind}. The time development of the averaged atom position $\bar{\mathbf R}_{j}(t)$ is adiabatically separated from its time-averaged displacement $\bar{\mathbf u}_{j}$ around $\bar{\mathbf R}_{j}(t)$.

This approximation implies that thermal equilibrium is maintained at each moment of the structural change, which means that the temperature of a glass is well-defined at each moment. 
Equation (\ref{eq:fundamental-relation1}) is now written as a function of time $t$,
\begin{equation}
U(t) = U_{t}\bm(T(t)\bm) + U_{s}\bm( \{ \bar{\mathbf R}_{j}(t) \} \bm),
\label{eq:ueq0t}
\end{equation}
where it is understood that the time $t$ in the equation is in the timescale much longer than $\tau_{t}$.
In fact, this approximation is what all previous studies assumed, regardless of whether they considered their treatment as equilibrium thermodynamics. Experimentalists can easily measure $T(t)$ as a function of time during a glass transition, 
and the change in the structure depends on the rate of temperature variation, $\gamma=dT/dt$. The glass transition is mechanically and thus thermodynamically a nonequilibrium process.
Now, the specific heat has both components corresponding to Eq.~(\ref{eq:fundamental-relation1}), namely, the phonon part $C_{t}$ and the structural part $C_{s}$:
\begin{equation}
C = C_{t} + 
 \frac{1}{\gamma} \sum_{j} \left( \frac{\partial U_{s}}{\partial \bar{\mathbf R}_{j}} \right) \frac{d \bar{\mathbf R}_{j}}{dt},
\label{eq:Ceq0t}
\end{equation}
where $C_{s}$ is given by
\begin{equation}
C_{s} = 
 \sum_{j} \left( \frac{\partial U_{s}}{\partial \bar{\mathbf R}_{j}} \right) \frac{d \bar{\mathbf R}_{j}}{dT}.
\label{eq:Cs}
\end{equation}

The entropy $S$ must be treated with care \cite{Broeck15}.
A simple analog to Eq.~(\ref{eq:ueq0t}), $S = S_{t}(T) + S_{s}( \{ \bar{\mathbf R}_{j} \} )$, does not hold for $S$. Whether entropy can exist for nonequilibrium states and how to define it if exists have long been notorious problems in statistical mechanics \cite{Penrose79,Lebowitz93,Mackey89,Uffink01,Grandy,Boksenbojm11,Lieb13,Swendsen17}. In Eq.~(\ref{eq:time-average}), the value of entropy depends on the period $t_{0}$ over which the time average is taken. 
Treating this major problem is beyond the scope of the present work. This difficulty, however, does not arise when the adiabatic approximation of the second kind holds.

Consider a simple example of diffusion of an interstitial atom. There are $N_{I}$ interstitial sites that a defect atom can occupy. Initially, the crystal is in the perfectly ordered state, which is specified by $\{\bar{\mathbf R}_{j}^{0} \}$. 
Suppose that the first atom, whose position is initially $\bar{\mathbf R}_{1}^{0}$, is displaced to a nearest interstitial site by electron irradiation. If $z$ nearest-neighbor interstitial sites are available, then $z$ different configurations are available for atom 1. Thereafter this atom repeats jumps at appropriate frequencies. For each jump, the atom occupies one of the $z$ nearest-neighbor interstitial sites. The jumping frequency (transition rate) is determined, of course, independent of the sampling time $t_{0}$.

The probability that the atom is found at site $i$ can be treated by taking ensembles. A different ensemble corresponds to a different trajectory. The number of atom configurations $\{ \bar{\mathbf R}_{j}^{K}(t) \}$, where $K$ is the configuration label, is well-defined at each moment. The dependence of the value $\bar{\mathbf R}_{j}$ on the integrating time $t_{0}$ has already been eliminated because the establishment of thermal equilibrium at each time guarantees a definite and unique value $\bar{\mathbf R}_{j}(t)$. The number $W_{K}(t)$ of atom configurations $\{ \bar{\mathbf R}_{j}^{K}(t) \}$ that occurs at a time $t$ can be counted by enumerating the ensembles. Upon normalizing by the number of ensembles, we obtain the distribution function $P_{K}(t)$ of atom configurations. The contribution of the structural part to entropy $S_{s}$ is given by
\begin{equation}
 S_{s}\left( \left\{ \bar{\mathbf R}_{j}^{(K)} (t) \right\} \right) = -\sum_{K} P_{K}(t) \ln P_{K}(t),
 \label{eq:SofSecondAdia}
\end{equation}
where $K$ is the collective label of the configuration.
This entropy part $S_{s}$ turns out to be the so-called configuration entropy.
The total entropy $S$ at time $t$ is thus obtained as
\begin{equation}
S(t) = S\left( U(t), \left\{ \bar{\mathbf R}_{j} (t) \right\} \right) =  
 S_{t}(U_{t}(t)) + S_{s}\left( \left\{ \bar{\mathbf R}_{j}^{(K)} (t) \right\} \right).
\label{eq:SofSecondAdia}
\end{equation}
Equation (\ref{eq:SofSecondAdia}) says that the entropy $S(t)$ of a solid at the present time is determined by the present distribution $\{ P_{K}(t) \}$ of atom configurations $ \{ \bar{\mathbf R}_{j}^{(K)} (t) \}$. The structural part $S_{s}$ carries information about the distribution of atom configurations $\{ P_{K}(t) \}$. By combining Eqs.~(\ref{eq:ueq0t}) and (\ref{eq:SofSecondAdia}), we obtain
\begin{equation}
U(t) =U(S(t), \{ \bar{\mathbf R}_{j}(t) \} ).
\label{eq:fundamental-relation-t}
\end{equation}
\begin{Theorem}[Properties of thermal equilibrium] \label{th:second}
The thermodynamic properties of a solid at the present time are uniquely determined solely by one of the representatives of the present atom configurations $\{ \bar{\mathbf R}_{j}^{(K)}(t) \}$, irrespective of the past history.
\end{Theorem}
Note that the configuration entropy can also exist in Eq.~(\ref{eq:fundamental-relation}). However, because constraints fix the atom positions $\{ \bar{\mathbf R}_{j} \}$, the freedom of changing atom configurations is frozen, and $\{ \bar{\mathbf R}_{j}^{(K)}(t) \}$ does not affect the thermodynamic properties of that solid. Using the wording of Ref.~\cite{StateVariable}, those coordinates are called {\it frozen coordinates}.

The maximum-entropy principle states that, for a fixed $U$, the equilibrium state is achieved when the total entropy $S$ obtains the maximum value \cite{Gibbs}. This requires
\begin{equation}
T_{s} = T_{t}, 
\label{eq:eqSandW}
\end{equation}
where $T_{t}=\partial U_{t}/\partial S_{t}$ is the phonon temperature and, therefore, the true temperature $T$, whereas $T_{s}$ is merely a parameter defined by $T_{s}=\partial U_{s}/\partial S_{s}$.

\paragraph{Internal variables.}
The existence of the FRE of Eq.~(\ref{eq:fundamental-relation-t}) is tacitly assumed in virtually all previous theories on nonequilibrium phenomena in solids. These theories assume (or take for granted) a specific and unique temperature. For the adiabatic approximation of the second kind, $T$ is well-defined. In this case, thermodynamic functions such as free energies are well-defined for each time of the change. However, additional TCs are required to describe the change in the structure. 

Traditionally, additional parameters called internal variables $Z$ (or order parameters) are introduced to describe nonequilibrium phenomena. The idea came from the analogy with the treatment of chemical reactions in gases, where the advancement of the reaction indicates how the reaction has progressed \cite{Prigogine54}. The advancement of the reaction is a dynamical variable describing the process, and therefore is not a TC: this is true for gas reactions for which two relaxation times $\tau_{t}$ and $\tau_{s}$ do not differ significantly. 

Various internal parameters are used in the glass literature \cite{Bouchbinder09-1,Gujrati10, Sciortino05,Nieuwenhuizen98a}. 
Among them, the fictive temperature $T^{\ast}$ is the most commonly used internal variable \cite{Davies53,Davies53a, Angell99, Rao02,Mysen-Richet05}. The fictive temperature is a parameter that reflects the instantaneous structure of a glass. Therefore, it is reasonable to assume that $T^{\ast}$ is a function of atom positions: $T^{\ast}=T^{\ast}( \{ \bar{\mathbf R}_{j} \} )$. 
As discussed in Sec.~\ref{sec:connect-thermo}, the property $Y$ of a material can serve as a TC of the material. Therefore, when the adiabatic approximation of the second kind holds, $T^{\ast}$ can be used as a TC. Theorem \ref{th:second} gives a rigorous base for the idea of fictive temperature. Similarly, enthalpy $H$ can be used as a TC. Later, $H$ is used to analyze the $C$-$T$ curve of glasses. One problem of the fictive temperature is to know what state this temperature refers. This problem is discussed in Sec.~\ref{sec:transition-state}.

\paragraph{Potential-energy landscape.}
The method of the potential-energy landscape (PEL) is widely used to study glass physics \cite{Stillinger95,Sastry98,Sciortino05}. Several features are common to the PEL method and the present theory.
First, in the PEL method, the system is described by the potential, which is a function of all atom positions, $V(\{ {\mathbf R}_{j} \})$. This is consistent with Theorem \ref{th:first}. The potential of a glass has many local minima called basins. Second, in the PEL method, the thermal average is obtained by using a partition function,
\begin{equation}
{\mathcal Q} = \sum_{i} \Omega(e_{i}) e^{-\beta e_{i}} 
 \int e^{-\beta \Delta V(\{ {\mathbf R}_{j} \}) } d \{ {\mathbf R}_{j} \} 
 = \sum_{i} \Omega(e_{i}) e^{-\beta f(e_{i},T,V)},
\label{eq:PEL-Z}
\end{equation}
with an appropriate normalization factor.
Here, $e_{i}$ is the energy minimum of basin $i$, $\Delta V(\{ {\mathbf R}_{j} \}) = V(\{ {\mathbf R}_{j} \})-e_{i}$, $f(e_{i},T,V)$ is the free energy of basin $i$, $\Omega(e_{i})$ is the number of basins of depth $e_{i}$, and $\beta$ is inverse temperature. The summation is over all basins. 
The use of the partition functions itself assumes thermal equilibrium, because only for this case does the parameter $\beta$ have sense. This treatment is consistent with the adiabatic approximation of the second kind in the present theory.
A difference between the PEL method and the present theory is that, in the PEL method, $V(\{ {\mathbf R}_{j} \})$ is treated as being independent of $T$, whereas it is not in the present theory. After integrating overall $\{ {\mathbf R}_{j} \}$, ${\mathcal Q}$ in the PEL method is a function only of $T$ and $V$. It follows that all thermodynamic quantities derived from ${\mathcal Q}$ are functions only of $T$ and $V$ (or $p$). However, it is unrealistic to consider that the potential does not change during the glass transition. The transition is defined by the change of structure.

To adapt this statistical treatment to the glass transition, the effective temperature $T_{\rm eff}$ is introduced as
\begin{equation}
 {\mathcal Q} = {\mathcal Q}(T_{\rm eff}, T, p) = \sum_{i} \Omega(e_{i}) e^{-\beta_{\rm eff} f(e_{i},T,p)},
\label{eq:PEL-noneq-Z}
\end{equation}
where $\beta_{\rm eff}$ is the inverse of $T_{\rm eff}$. Only when equilibrium is established do we have $T_{\rm eff}=T$. Although the phenomenological introduction of $\beta_{\rm eff}$ leads to useful results for many problems of glasses, this parameter is not justified on physical grounds \cite{Sciortino05}.

\section{Glass state and glass transition}
\label{sec:eq-state-glass}

\subsection{The glass-transition temperature in $C$-$T$ curve}
\label{sec:glass-transition}

\begin{figure}[htbp]
\centering
\includegraphics[width=80 mm, bb=0 0 500 340]{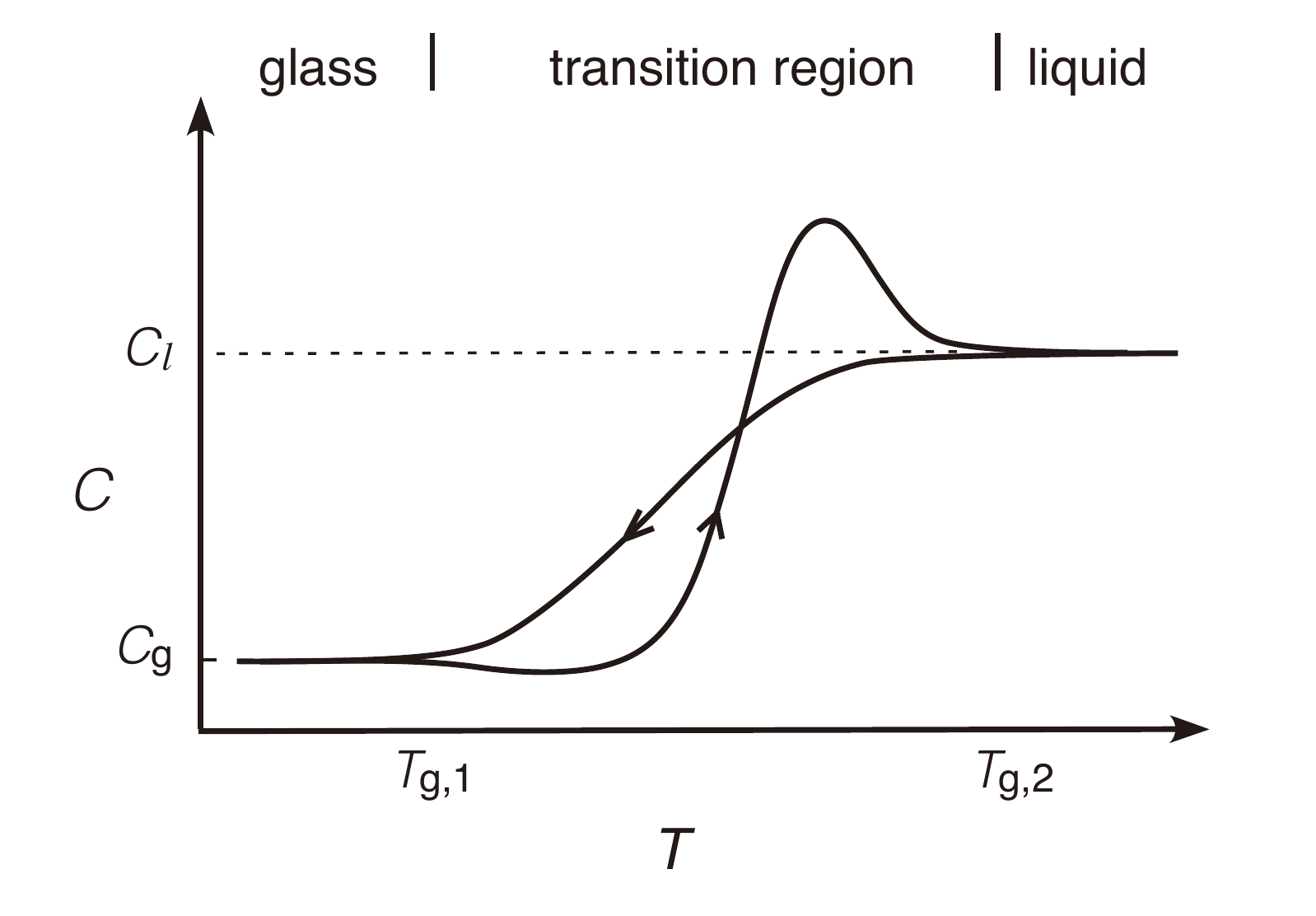} 
\caption{Glass transition observed in the specific heat $C$. The cooling and heating processes are indicated by arrows.
} \label{fig:CTcurve}
\end{figure}

We now discuss the glass transition, which occurs over a narrow range of temperature in the $C$-$T$ curve. An example is shown in Fig.~\ref{fig:CTcurve}. 
Generally, the specific heat $C_l$ of the liquid state is greater than that of the solid state $C_{g}$.
When a supercooled liquid of a glass is cooled, the specific heat of the liquid decreases rapidly at the temperature $T_{g,2}$. This rapid decrease ceases at $T_{g,1}$, which indicates the termination of the solidification process. Below $T_{g,1}$, the glass substance is in the solid state: the term ``glass state" is reserved herein to refer to this state. 
In the range $T_{g,1} < T < T_{g,2}$, the state of the glass substance is referred to as the {\it transition state}.
Although the definition of the glass-transition temperature $T_{g}$ within the width $\Delta T_{g} = T_{g,2} - T_{g,1}$ is somewhat arbitrary \cite{Moynihan76}, this does not matter for the present study.

The $C$-$T$ curve depends on the heating and cooling rate, so the process changes $T_{g}$. This fact leads some researchers not to regard $T_{g}$ as an inherent property of glasses \cite{Adam65,Gupta07,Hentschel08,Berthier11}. However, the change in $T_{g}$ is only slight: for glycerol, a one-week prolongation of the transition state altered the $T_{g}$ by only several degrees \cite{Oblad37}. $T_{g}$ has a physical significance in the same degree as the melting temperature $T_{m}$ does in crystalline materials. 
In the glass literature, the frequently used term {\it kinetically frozen} hinders the generality of the kinetic nature of solidification. Any solidification is a kinetically frozen process, in which the potential energy overcomes the kinetic energy.
In crystals, the melting temperature $T_{m}$ is determined by the balance between the potential part $\Delta H_{m}$ and the kinetic part $T_{m} \Delta S_{m}$ of the free energy, namely, $\Delta H_{m}=T_{m} \Delta S_{m}$, where $\Delta H_{m}$ is the enthalpy of fusion and $\Delta S_{m}$ is the entropy of fusion. The latter is essentially the configurational entropy $S_{c}$ of the liquid.
For glasses, there is a well-known formula by Adam and Gibbs for modeling the temperature dependence of viscosity $\eta$ in the vicinity of $T_{g}$,
\begin{equation}
\eta=A' \exp \left( \frac{B' \Delta \mu}{TS_{c}} \right),
\label{eq:AGeq}
\end{equation}
where $A'$ and $B'$ are constants and $\Delta \mu$ is the potential barrier for atom rearrangement \cite{Adam65}. Considering that the glass transition occurs near $\eta \approx 10^{13}$ Poise, we see that Eq.~(\ref{eq:AGeq}) is looked upon as giving the condition of the glass transition. By choosing appropriate values $A'$ and $B'$, we have $\Delta \mu=T_{g} S_{c}$, which has the same form as the melting temperature of crystals.

Hence, the dependence of $T_{m}$ on the preparation conditions is not specific to glasses alone. 
The theory of crystal growth tells us that supercooling is indispensable for complete crystallization, no matter how small the effect is \cite{Porter-PT-metals}; otherwise, the crystallization process would require infinite time. The transition pressure and temperature of the graphite-diamond transition have been established \cite{Bundy85}. However, deviations from these values are often observed owing to its kinetics; now, diamond can be synthesized even at low-pressure processes \cite{Spitsyn81,Kamo83}. Our experience shows that the kinetics of reaction agents drastically the phase boundary of boron crystals \cite{Uemura19}.

The curve in Fig.~\ref{fig:CTcurve} is produced by a heat cycle starting from the liquid state, cooling to the glass state, and then heating to restore the liquid state. After completion of a cycle, the enthalpy $H$ returns to its initial value, so the following integral is satisfied:
\begin{equation}
\oint C dT = 0.
\label{eq:circ-heat0}
\end{equation}
Generally, this closed relation is not guaranteed when a heat cycle consists of starting from the glass state, heating to the liquid state, and cooling to the glass state. This means that the final state of the heat cycle differs from the initial state. Despite this, the specific heat $C_{g}$ of the glass state remains constant.

\subsection{Glass state}
\label{sec:eq-state-glass}

\subsubsection{Equilibrium nature of glass state}
\label{sec:sub:eq-state-glass}

Presently, the conventional view holds that the glass state is far out of equilibrium. However, this argument overlooks the role of constraints described in Sec~\ref{sec:TDfundaments}. To understand this, consider water in a container, which is in a stable equilibrium, irrespective of the elevation the container is held at. However, if a hole punctures the container, the water immediately flows out to the ground. The state of water in the elevated position is quite unstable. The constraint ({\it i.e.}, the container) thus renders stable an otherwise unstable state. In a similar manner, the energy barriers built around the atoms render stable an unstable structure of a glass.
Based on Theorem \ref{th:first}, the glass state is an equilibrium state because the position of all atoms is fixed. 
Furthermore, multiple reasons explain why the glass state is an equilibrium state, as detailed below.

To begin with, viewing the glass state as an equilibrium state is compatible with the zeroth law of thermodynamics, which defines temperature in terms of an equilibrium between two systems being guaranteed when the two systems have the same temperature. Of course, the temperature of glasses can be measured by the usual methods, similar to other materials. No change occurs when a glass at temperature $T$ comes in thermal contact with another system at the same $T$.
This is by no means a trivial matter. If glass were not an equilibrium state, work could be extracted from it without changing other conditions. This is tantamount to extracting work from only one heat bath, which contradicts the second law of thermodynamics.

Second, note that the belief that glass will crystallize if given sufficient time is merely speculation about future events. In mechanics, if a stone rests on a slope, there is general agreement that this state is an equilibrium state. The interpretation is that equilibrium is established by the frictional force canceling the gravitational force exerted on the stone. However, given sufficient time, the stone may move downward. If we were to regard the stone to be in a nonequilibrium state based on its future state, nothing could be claimed with certainty. We can thus speak of equilibrium only within the constraints that are valid at the present time. 

Third, the view of glass as a nonequilibrium state conflicts with observed facts. A current observation is that glasses can maintain their structure for over one million years \cite{Berthier16}, whereas over a similar time span, most metals degrade due to corrosion, oxidation, and other effects. 

Fourth, glass is stable from the viewpoint of solid-state theory. The phonon spectra of oxide glasses have no soft mode \cite{Galeener83,Sen77,Stich91}, which implies that the glass states are dynamically stable, which is consistent with their high melting points. Of course, they are elastically stable, too.

\subsubsection{Reconciliation with the third law of thermodynamics}
The belief that the glass state is a nonequilibrium state arose for historical reasons. Nonvanishing entropy at $T$ close to zero was found in glass materials in the early 20th century \cite{Fowler-Guggenheim, Wilson}, which contradicts the third law of thermodynamics. Researchers resolved this problem by deeming that the glass state was out of equilibrium. They considered that thermodynamics is a theory for equilibrium states only, and that glasses are therefore not subject to the laws of thermodynamics (see, e.g., Ref. \cite{Wilks}, p.~63).
This is not true. The definition of equilibrium shows that work, which is the quantity of interest in many applications, can only be obtained when a system is out of equilibrium.
The issue of the third law of thermodynamics is also deeply related to the definition of equilibrium and was solved in previous work \cite{Shirai18-res}. Here, only the main results are summarized, and the reader is referred to Ref.~\cite{Shirai18-res} for more details.

The statement of the third law of thermodynamics is that two materials have a common origin in entropy $S$ at $T=0$ when the dimensions of the state spaces of the two materials are the same. The absolute value of the entropy $S_{0}$ at the origin is irrelevant. When the state spaces have different dimensions, their origins generally differ by a finite amount, and this difference appears as the residual entropy. The TCs that are not common to both materials are the frozen coordinates, which do not influence the thermodynamic properties of the materials (Theorem \ref{th:first}). Note that we {\em can} take $S_{0}=0$ provided only this material is investigated.
Thus, the entropy of a particular sample of glass {\em can} be taken as $S_{0}=0$ at $T=0$, because the particular sample occupies only one atom configuration $K$. 
When we compare the entropy of a glass state with that of a crystal state, their respective origins must be adjusted. This is possible only by removing the constraints that maintain the frozen coordinates (Theorem \ref{th:second}). At this moment, the glass turns out to be in a nonequilibrium state, which must change to the equilibrium state, namely, the crystal state. Upon reaching this final state, the entropy difference vanishes, and the third law of thermodynamics is recovered. Thus, the problem of residual entropy does not conflict with the third law of thermodynamics.

\subsubsection{Randomness and order parameter}
A misconception of randomness makes understanding the nature of glass more difficult.
Unfortunately, even now the literature mistakes the absence of periodicity for disorder, despite the caution urged by several authors \cite{Ben-Naim,Rosenkrantz83, Grandy,Denbigh89,Styer00}. 
A lack of periodicity does not necessarily imply disorder. The structure of DNA has no periodicity, but yet the entropy is very small. Our town has geometrically no regularity. Despite this, a postman can correctly deliver postal materials to any address, if he has a town map.
The distinction between order and disorder is made by whether information is missing.
As discussed in Sec.~\ref{sec:equilibrium}, in spite of the apparent randomness, the information of atom position in a glass is not destroyed by averaging over time. This is reflected in the specific heat of a glass, which is almost the same as that of its crystal phase \cite{Gibson23,Simon26,Park28,Chang72}. This means that the entropies of the two phases are almost the same, aside from the residual entropy, which is deactivated because of the frozen coordinates.

Based on the entropy, the degree of order of the glass state is almost the same as that of the crystal, provided the frozen coordinates (the configuration entropy, in this case) are disregarded. In this sense, the TC is akin to an order parameter. 
Since the notion of order parameter was introduced for describing the glass transition by the pioneering work of Davies and Jones \cite{Davies53}, discussion has continued on the nature of the order parameter of glasses \cite{Berthier16,Gupta76,Lesikar80,Franz97,Xia00,Berthier11,Charbonneau14}. The elusive notion of order in disordered materials has yielded various definitions of the order parameter.
We now see that the order parameter is no more than a TC. Any property $Z$ that is characterized by the structure, $Z=Z(\{ \bar{\mathbf R}_{j} \})$, can be used to indicate the order of glasses.

\subsection{Transition state of glass}
\label{sec:transition-state}
An important issue regarding the nature of the glass state lies in the interpretation of the transition region of a $C$-$T$ curve. 
Glass transitions are nonequilibrium phenomena because, by definition, transitions are time dependent. Despite this, thermodynamics methods are not invalidated if the adiabatic approximation of the second kind holds, as discussed in Sec.~\ref{sec:second-adiabatic}. This is the case for the transition region. We can treat the structural and thermal parts of the state separately and treat the thermal part as being in thermal equilibrium. The structural part conveys irreversibility because it involves relaxation processes, which give rise to hysteresis. A hysteresis appears in the heat cycle of the glass transition in Fig.~\ref{fig:CTcurve}. Thus, glass transitions are generally irreversible processes, unless special care is taken.

\begin{figure}[htbp]
\centering
\includegraphics[width=80 mm, bb=0 0 560 460]{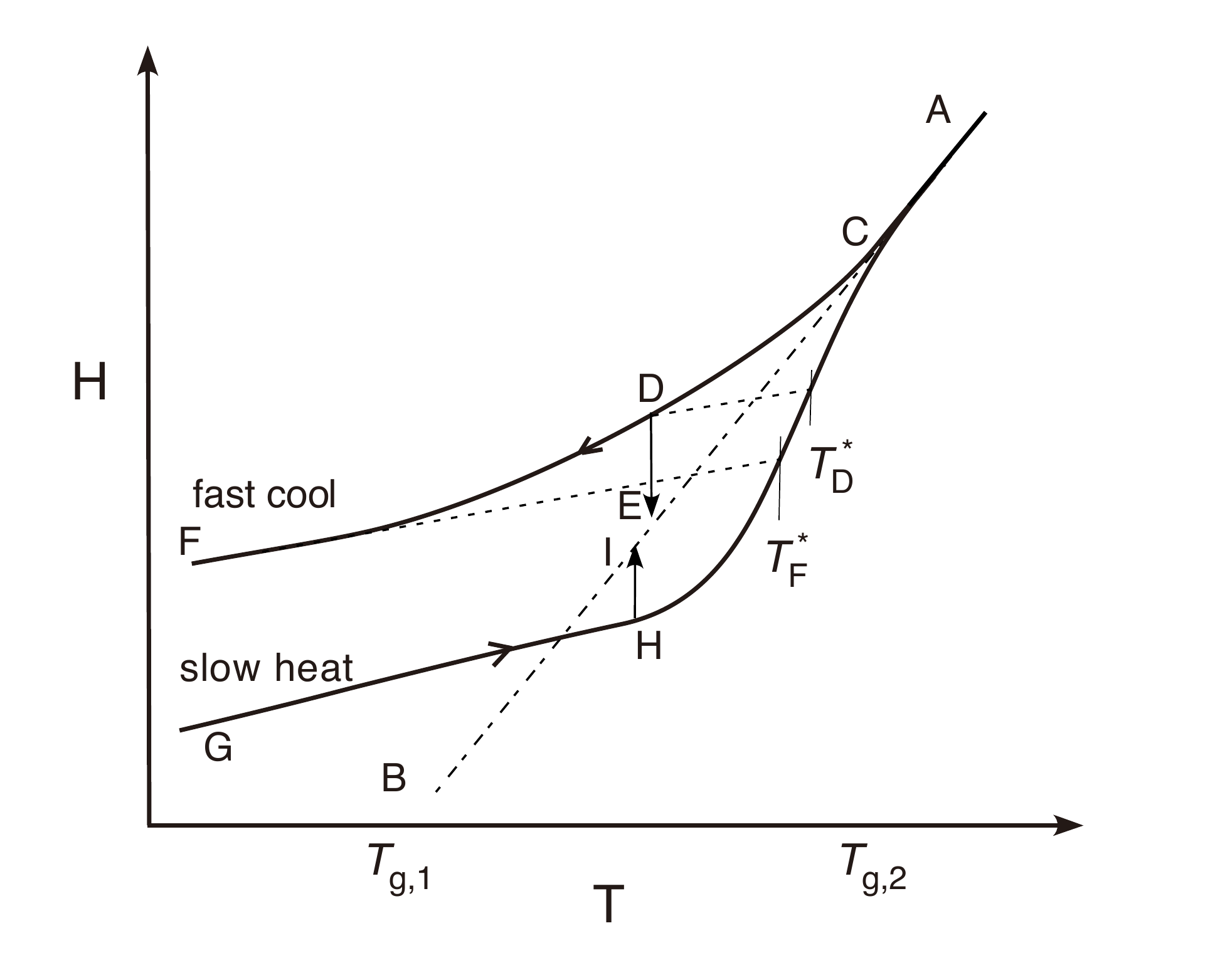} 
\caption{Enthalpy $H$ versus temperature $T$ in the glass transition. Cooling and heating processes are indicated by arrows. Note that two curves for cooling and heating processes are obtained for the two samples, which had different heat treatments. 
} \label{fig:HTcurve}
\end{figure}
The $C$-$T$ curve is affected by the preparation conditions, and in particular by altering the cooling and heating rate $\gamma$. An interesting question about the glass transition is toward what state does the glass transition? It is generally considered that glass finally transitions to the supercooled-liquid state if given sufficient time. This is explained by using the $H$-$T$ curve of Fig.~\ref{fig:HTcurve}, which is commonly used in the literature \cite{Davies53a,Ritland54, Narayanaswamy71,Moynihan74}. This curve is obtained from a $C$-$T$ curve by integrating the measured values of $C$ with respect to $T$. The following explanation is based on the study of Davies and Jones \cite{Davies53a,Davies53}.

A glass state is obtained by cooling a supercooled-liquid state of a glass substance. The enthalpy $H$ of supercooled liquid is indicated by the dashed-dotted line AB. Actually, the low-temperature part of AB is a linear extrapolation of the high-temperature part of the $H$-$T$ curve. The supercooled-liquid state is considered an equilibrium state over the entire range of $T$. The {\it ideal glass} should appear on this extrapolated line. When cooled to a temperature slightly below $T_{g,2}$, the enthalpy $H$ of the glass substance begins to deviate from the equilibrium line AB. As $T$ decreases further, the deviation increases. At a point D ($T_{g,1} < T_{\rm D} < T_{g,2}$), we stop the cooling and detach the system from the heat bath. A relaxation process begins toward the equilibrium state AB. If we wait sufficiently long, the supercooled-liquid state will be recovered. This is the view that most researchers envisage.

However, no evidence exists to indicate that the stability of the supercooled-liquid state at $T>T_{g,2}$ can be extrapolated to lower temperatures. This is merely a speculation. An excuse for the lack of evidence is that the time required to reach equilibrium is extremely long for accessing by experiment. 
The reason for the stability of the supercooled liquid is thus largely based on the concept of ergodicity, which claims that every atom of a glass substance visits every position inside the glass. However, the high viscosity of glass prevents the atoms from visiting everywhere except the vicinity of their initial position. This reasoning has pervaded the research of glass for over a century. 
However, this is a one-sided argument, which overtakes the role of ergodicity. Thermodynamic equilibrium is established by two competitive forces: the active tendency and the passive resistance in the words of Gibbs (\cite{Gibbs}, p.~58). The former is driven by ergodicity, whereas the latter is caused by constraints, to use the current terminology. Crystallization is the consequence of overcoming the constraint over ergodicity.

The stability of supercooled liquid at $T<T_{g,2}$ is incompatible with the thermodynamic stability of a material, namely, the fact that specific heat is a positive quantity (see, e.g., Ref.~\cite{Callen}, p.~206). The fact that specific heat is positive ensures that the free energy $G(T)$ is concave with respect to $T$, so that, below the transition temperature, the low-temperature phase must be stable relative to the high-temperature phase. The appendix shows compelling evidence in support of the concave curve of $G(T)$. 
Therefore, the idea that the transition state approaches the supercooled-liquid state has never been proven. For the same reason, the transition state cannot be a purely solid glass state in a range $T_{g,1} < T < T_{g,2}$. 
The transition state is an inhomogeneous mixture of the solid and liquid phases, as is usual for crystallization, where nucleation of solid occurs. The nucleation of the glass state may occur in the transition region \cite{Kelton91}. Novel notions, such as dynamical heterogeneity (for example, see Ref.~\cite{Rao02}, p.~100), may be interpreted as this nucleation, although the author is not sure. 

Because glasses have many local minima, there is no unique equilibrium state. Depending on the initial configuration $\{ \bar{\mathbf R}_{j}^{K} \}$, the glass substance finds a local minimum near the initial configuration. For practical purposes, we choose as equilibrium states of glass those that are obtained by cooling at a rate as slow as experimentally possible. These states correspond to the point on the $C$-$T$ curve that is obtained by the slowest cooling. This assumption is used in the next section.

\section{Analysis of a $C$-$T$ curve in the heating process}
Let us apply the foregoing theory to the glass transition in a $C$-$T$ curve. The $C$-$T$ curve in the transition region depends on the rate of change in temperature, $\gamma$, and this section describes this dependence by using the present theory. The result has already been treated by many authors in the traditional manner \cite{Moynihan74,Moynihan76,Debolt76,Hodge83,Hutchinson76,Kovacs79}. Thus, the results of the calculation are not the subject; instead, the logical connection behind the results provides the main interest.

\subsection{Theoretical model}

\subsubsection{Heating process of glass}
\begin{figure}[htbp]
\centering
\includegraphics[width=120 mm, bb=0 0 1100 620]{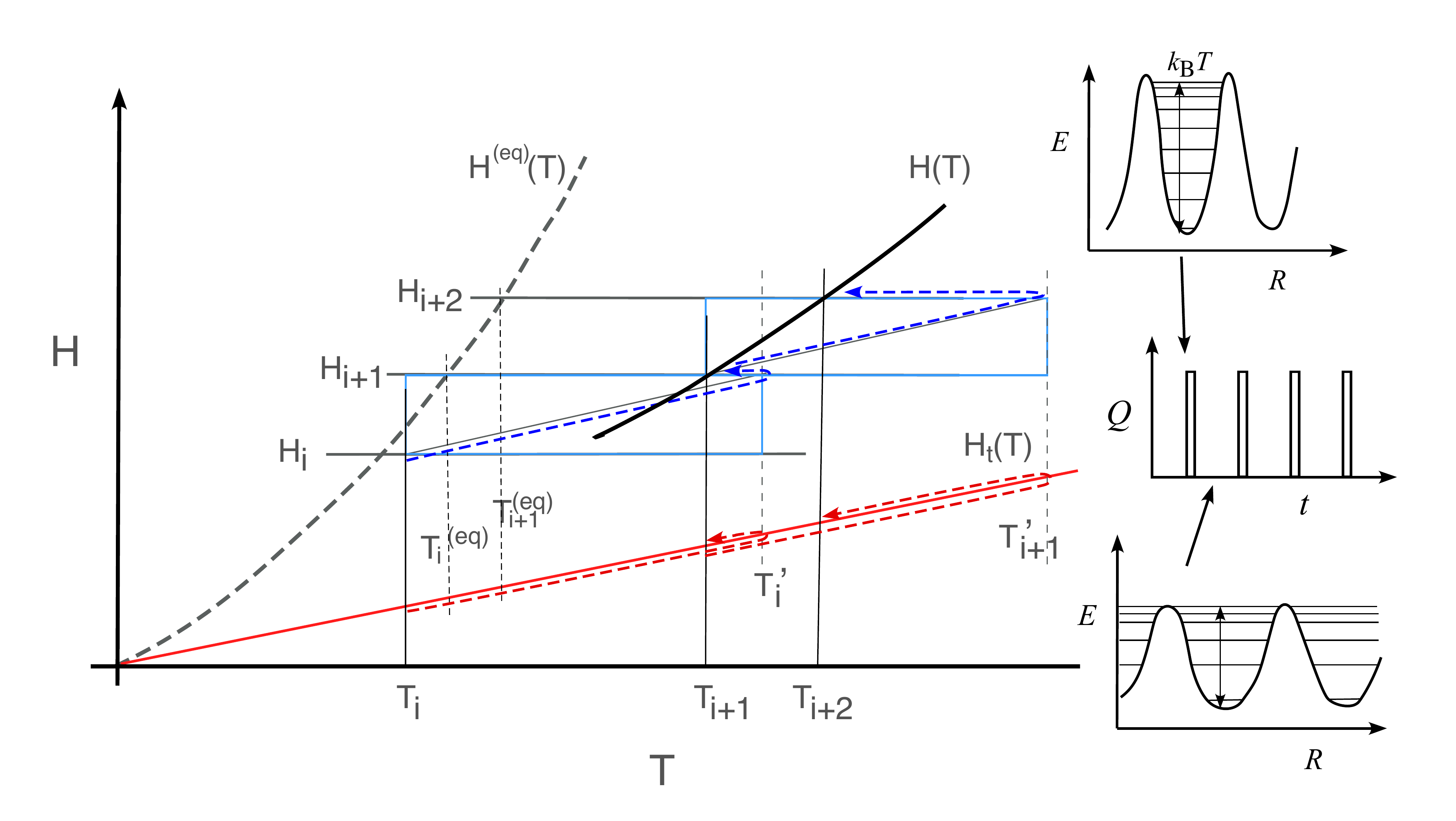} 
\caption{Heating process in an $H$-$T$ curve, together with the $T$ dependence of the phonon part, $H_{t}$, and that of the enthalpy of the equilibrium state, $H^{\rm (eq)}$. The blue lines indicate the change in total enthalpy $H$, and the red lines indicate the change in the phonon part $H_{t}$.
} \label{fig:adia-CT}
\end{figure}

In this subsection, the glass substance is referred to simply as a system. 
As discussed in Sec.~\ref{sec:connect-thermo}, the enthalpy $H$ can be used as a TC, in addition to $T$. This choice of TC may be the best choice because the dependence of $H$ on the structure $\{ \bar{\mathbf R}_{j} \}$ is rigorously based in DFT. 
The state of the system is thus expressed in a two-dimensional state space ($T, H$), as shown in Fig.~\ref{fig:adia-CT}. Only when the system is in thermodynamic equilibrium is there a unique relationship between them: $H(T)=H^{\rm (eq)}(T)$. 
The enthalpy $H$ of a system is decomposed into the phonon part $H_{t}$ and the structural part $H_{s}$, as in Eq.~(\ref{eq:ueq0t}).

Consider a heating process with a constant heating rate $\gamma>0$. The heating is accomplished by consecutive injections of heat pulses with period $t_{p}$. Except for the injection of heat pulses, the system is adiabatically isolated from the surroundings.
Although most previous studies considered isothermal processes \cite{Moynihan74,Moynihan76, Debolt76,Hodge83}, adiabatic processes are desirable because the separation between $H_{t}$ and $H_{s}$ clarifies the mechanism. The relaxation occurs by adjusting the portions between $H_{t}$ and $H_{s}$, while the total $H$ is maintained constant. This is consistent with the expression of the second law of thermodynamics: the state of a system changes to achieve the maximum entropy while the internal energy remains fixed \cite{Gibbs}. In this respect, the present model is the same as that used by Davies and Jones \cite{Davies53}. For the adiabatic approximation of the second kind, the temperature $T$ of the system is understood to be the temperature of the phonon subsystem, which is always in equilibrium.

A pulse of heat $\Delta Q_{i}$ is injected into the system at $t=t_{i}$. Just before the heat injection, the temperature of the system is $T(t_{i}-)=T_{i}$.
The duration $t_{w}$ of the pulse is very short, but much longer than $\tau_{t}$, so that the phonon subsystem can immediately follow the heat input and equilibrate. 
At this moment, the enthalpy of the system changes with the phonon part $H_{t}$ only; namely, $\Delta H_{t}(t=t_{i}+) = \Delta Q_{i}$, while $\Delta H_{s}(t=t_{i}+)=0$ because of the slow response of the structure. 
Using the adiabatic approximation of the second kind, the increase in $T$ is given by $\Delta T(t=t_{i}+) = \Delta Q_{i}/ C_{t}$, so the temperature of the system becomes $T(t=t_{i+}) = T_{i}' = T_{i}+ \Delta Q_{i}/ C_{t}$.

Just after the heat injection, the system undergoes an adiabatic change toward the thermodynamic equilibrium state. Because the process is adiabatic, the total enthalpy of the system $H$ is preserved, so $H=H_{i+1} \equiv H_{i}+\Delta Q_{i}$. The total enthalpy $H(T)$ moves along the blue curve in Fig.~\ref{fig:adia-CT}. Only the portions between $H_{t}$ and $H_{s}$ vary with time. The second law of thermodynamics dictates that the state changes so as to maximize the entropy of the system under the constraint of a fixed $H$. 
The rate of change of $H_{s}(t)$ is proportional to the deviation $\Delta H_{s}$ from the equilibrium value $\Delta H_{s}^{\rm (eq)}$,
\begin{equation}
\frac{d H_{s}}{d t}  = -\frac{\Delta H_{s}}{\tau_{s}},
\label{eq:total-H}
\end{equation}
where the proportionality constant $\tau_{s}$ is the structural relaxation time. 
Thus, $H_{t}(t)$ is
\begin{equation}
H_{t}(t) = \left\{ H_{t}(T'_{i}) -H_{t}(T_{i}^{\rm (eq)}) \right\} e^{-t/\tau_{s}} + H_{t}(T_{i}^{\rm (eq)}),
\label{eq:relax-H}
\end{equation}
where $H_{t}(T_{i}^{\rm (eq)})$ is the enthalpy of the phonon part when the thermodynamic equilibrium is reached for step $i$. 
The critical point that differs from previous studies is that the equilibrium state into which the system transitions is the state obtained by cooling as slow as possible, as described in the previous section. The adiabatic change causes a change in $T$ given by
\begin{equation}
T(t) = (T'_{i}-T_{i}^{\rm (eq)}) e^{-t/\tau_{s}} + T_{i}^{\rm (eq)}.
\label{eq:relax-T}
\end{equation}
For a given $H_{i+1}$, only a single thermodynamic equilibrium state satisfies $H(T_{i}^{\rm (eq)}) = H_{i+1}$.

Before reaching the equilibrium state, the next heat pulse $\Delta Q_{i+1}$ is injected at $t=t_{i+1}=t_{i}+t_{p}$. The final temperature after heat pulse $i$ is then $T_{i+1} = T(t=t_{i}+t_{p})$ in Eq.~(\ref{eq:relax-T}).
The specific heat at this step is
\begin{equation}
C_{i} = \frac{\Delta Q_{i}}{\Delta T_{i}},
\label{eq:CatIth}
\end{equation}
where $\Delta T_{i}=T_{i+1}-T_{i}$. The next step begins at $T_{i+1}$.
A constant rate $\gamma$ is achieved by taking $\Delta Q_{i+1}$ to be $C_{i} \gamma t_{p}$. This process continues until $T$ becomes greater than $T_{g,2}$.

\begin{figure}[htbp]
\centering
\includegraphics[width=80 mm, bb=0 0 610 360]{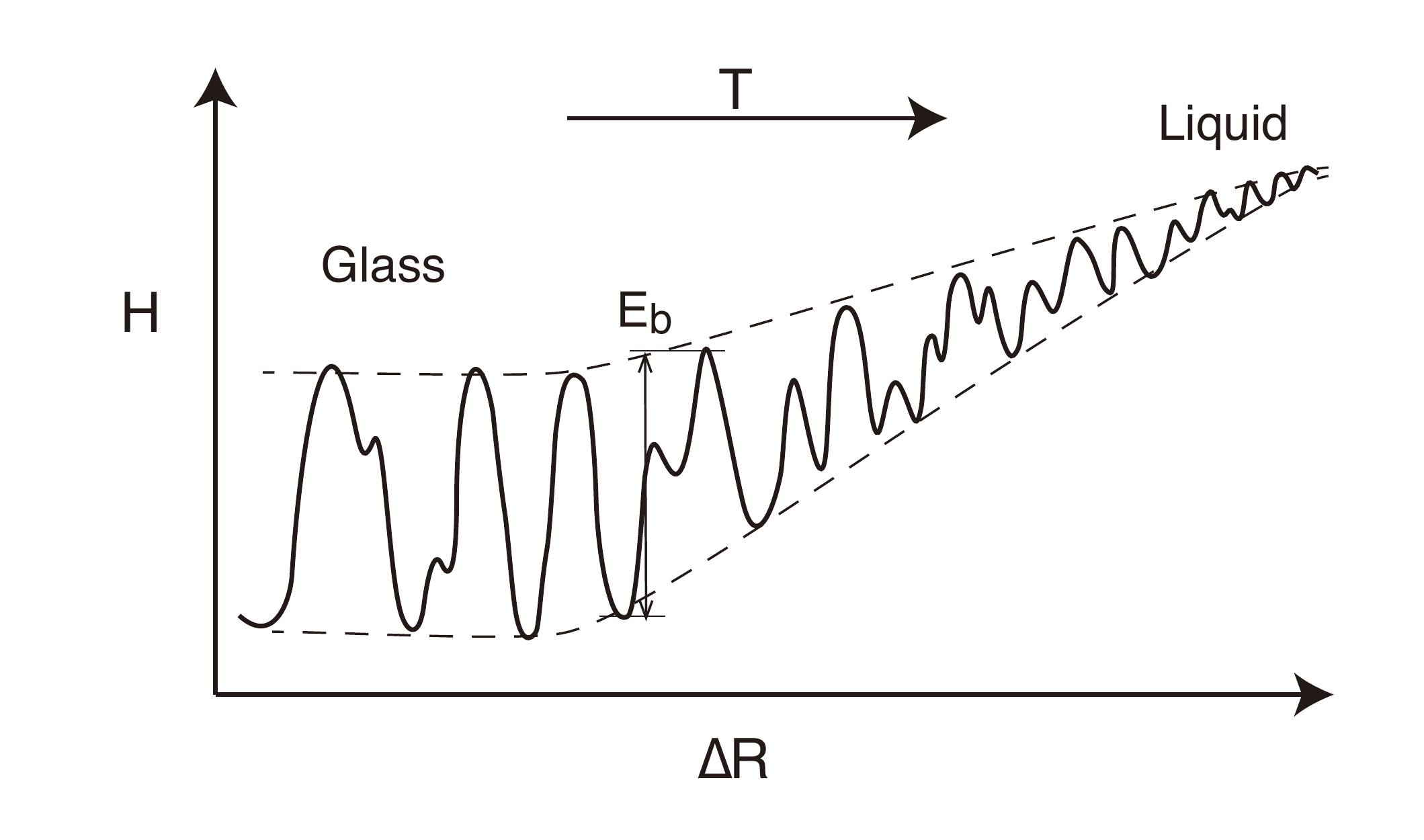} 
\caption{Energy landscape for glass transition. Enormous basins are apparent. The enthalpy $H$ is expressed as a function of the atom coordinate $R$. The abscissa qualitatively represents temperature, too. 
} \label{fig:landscape}
\end{figure}

\subsubsection{Activation energy}
Another important feature of the present model lies in the determination of the relaxation time $\tau_{s}$.
In typical relaxation phenomena, such as diffusion in solids, the relaxation time $\tau$ obeys the activation law 
\begin{equation}
\frac{1}{\tau} = A \exp \left( -\frac{E_{a}}{k_{\rm B}T} \right),
\label{eq:relax-tau}
\end{equation}
where $E_{a}$ is the activation energy and $A$ is an appropriate constant. Normally, $E_{a}$ is considered to be independent of $T$. 
In the glass transition, the assumption of constant $E_{a}$ leads to poor agreement with experiment. The response of glasses to external perturbations have been known since the early days to be nonlinear \cite{Lillie36} and non-exponential \cite{ Ritland56}. 
To treat these properties, Narayanaswamy {\it et al.} replaced the factor $e^{-t/\tau_{s}}$ in Eq.~(\ref{eq:relax-H}) with a time-dependent response function $\phi(t,t')$ \cite{Gardon70,Narayanaswamy71,Debolt76}. They took the thermal history into account through the form $\tau=\tau(T, T^{\ast})$, and the structural dependence is accounted for by introducing the fictive temperature $T^{\ast}$, giving
\begin{equation}
\phi(t-t') = \exp \left[ -\int_{t'}^{t} \frac{dt'}{\tau(T, T^{\ast} )} \right].
\label{eq:relax-tau-fictive}
\end{equation}
By using the reduced time $\zeta = \int dt' /\tau$, the expression of Eq.~(\ref{eq:relax-tau-fictive}) is rewritten as $\phi(t-t')=\exp (-\zeta) $. In accordance with the custom followed in the study of dielectric relaxation, the response function takes the form
\begin{equation}
\phi(t-t') = \exp (-\zeta^{\beta}),
\label{eq:relax-beta}
\end{equation}
where $\beta$ is a constant, with $0<\beta \le 1$. 

There remains the task of determining $\tau(T, T^{\ast} )$. Moynihan {\it et al.} did this by replacing $E_{a}$ in Eq.~(\ref{eq:relax-tau}) by
\begin{equation}
E_{a} = \Delta h^{\star} T \left\{ x \frac{1}{T}+(1-x)\frac{1}{T^{\star}} \right\},
\label{eq:activation-H}
\end{equation}
where $x$ and $\Delta h^{\star}$ are fitting parameters \cite{Tool46,Gardon70,Narayanaswamy71,Moynihan74,Moynihan76,Moynihan76a}. This equation is known as the Tool-Narayanaswamy-Moynihan (TNM) formula. In this case, $\Delta h^{\star}$ is the activation energy instead of $E_{a}$. 
In this manner, the calculation of the relaxation process became tractable, and this method using the four parameters $A$, $\Delta h^{\star}$, $x$, and $\beta$ became a standard approach for today's glass research. However, by examining the use of this four-parameter model for a wide range of glasses, Hodge concludes that no single set of these parameters provides satisfactory fitting for glasses with various histories \cite{Hodge94}.

In contrast with the previous studies, we retain the original exponential decay in Eq.~(\ref{eq:relax-H}), and $E_{a}$ in Eq.~(\ref{eq:relax-tau}) is maintained as the activation energy. Instead, we assume that $E_{a}$ and therefore $\tau_{s}$ depend on the structure of the solid at the present time, according to Theorem \ref{th:second}. The activation energy for the glass transition is in a wider sense the energy barriers for atom migration and trapping. The behavior of atoms of glass in the transition region is expressed by an energy landscape, which is shown schematically in Fig. \ref{fig:landscape}. As the glass cools, the energy barrier $E_{b}$ develops around atoms and impedes their migration. The calculation of its structural dependence, $E_{b}=E_{b}(\{ \bar{\mathbf R}_{j} \})$, is very complicated; however the details of the dependence are usually not overly significant. As $H$ is a function of $\{ \bar{\mathbf R}_{j} \}$, it can be simplified by the form $E_{b}=E_{b}(H)$.
It is natural to consider that $E_{b}$ increases as $H$ decreases from the enthalpy of the liquid $H_{l,0}$ at $T=T_{g,2}$ \cite{Debenedetti01}. To first-order, we assume a linear dependence on $H$:
\begin{equation}
E_{b}(H) = 
\left\{
\begin{array}{ll}
0  & (H>H_{l,0}) \\
b (H_{l}-H) \ \ \ & (H_{g,0}<H<H_{l,0}) \\
E_{b0}  & (H<H_{g}), \\
\end{array}
\right.
\label{eq:Eb-Hdep}
\end{equation}
where $b=E_{b0}/\Delta H_{lg}$, $\Delta H_{lg}=H_{l,0}-H_{g,0}$, and $H_{g,0}$ is the enthalpy of the glass at $T=T_{g,1}$. Setting $E_{b}(H) = 0$ for $H>H_{l,0}$ is, of course, an approximation for a small value $E_{b}$ in that region. As usual in chemical reaction, the energy barrier varies depending on the direction of reaction, and hence $E_{b}$ is different for cooling and heating processes. Here, this difference is ignored for the sake of simplicity.
In this manner, the complicated task of tracing the thermal history is tremendously simplified. In principle, the number of $\tau_{j}$ is $N_{\rm at}$. As usual, we proceed as far as possible with a simple model that uses a single relaxation time.

\subsection{Simulation result}
\label{sec:simulation}

\begin{figure}[htbp]
\centering
\includegraphics[width=85 mm, bb=0 0 400 260]{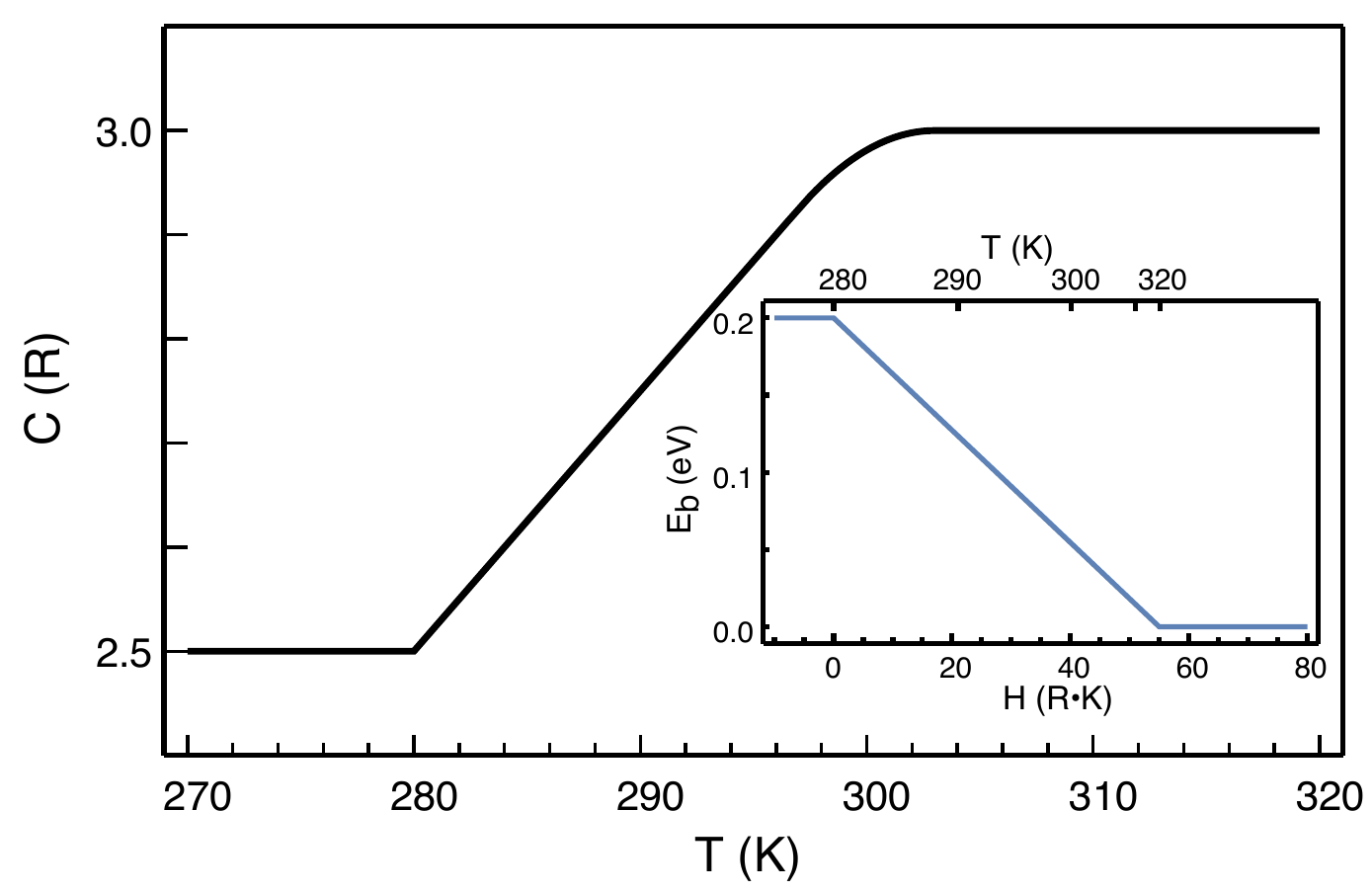} 
\caption{Specific heat $C$ of a hypothetical glass in equilibrium. $C$ is presented in units of the universal gas constant $R$. The glass transition occurs in a temperature range between $T_{g,1}=280$~K and $T_{g,2}=300$~K. The inset shows the assumed energy barrier $E_{b}$ for the structural relaxation as a function of $H$.}
\label{fig:c-T}
\end{figure}

\subsubsection{Dependence on cooling or heating rate}
\label{sec:heating}
We now present the results of the simulation of the glass transition. The specific heat $C^{\rm (eq)}$ of the equilibrium state is prescribed for a hypothetical glass. Given the energy barrier $E_{b}$ for the structural relaxation in the form of Eq.~(\ref{eq:Eb-Hdep}), we calculate how the specific heat $C$ depends on the cooling or heating rate $\gamma$. 
Figure \ref{fig:c-T} shows the assumed $C^{\rm (eq)}$-$T$ curve for the hypothetical glass in the equilibrium state, which is obtained by cooling at the slowest possible rate: $\gamma \rightarrow 0$. 
The specific heat $C^{\rm (eq)}(T)$ is assumed to be linear in temperature in the transition region. Around $T_{g,2}$, the curve for $C^{\rm (eq)}$ connects smoothly to $C_{l}$. The curve for $C^{\rm (eq)}$ as a function of $T$ fixes the values of $H$ at $H_{l,0}^{\rm (eq)}=54.9$~eV by taking the origin of $H$ to be $H_{g,0}^{\rm (eq)}=0$ at $T_{g,1}$. The inset of Fig.~\ref{fig:c-T} shows the energy barrier $E_{b}$ as a function of $H$. 
Because $E_{b}$ depends exponentially on $\tau_{s}$, the value $E_{b0}$ is sensible for obtaining reasonable results. We use $E_{b0}=0.2$~eV in the following simulations. Using twice or half of this value would force us to use unrealistic values of $\gamma$ to obtain a reasonable dependence of the $C$-$T$ curve on $\gamma$. In this study, we assume a range of $\gamma$ from 0.1 to 10 K/s.

\begin{figure}[htbp]
\centering
\includegraphics[width=80 mm, bb=0 0 420 280]{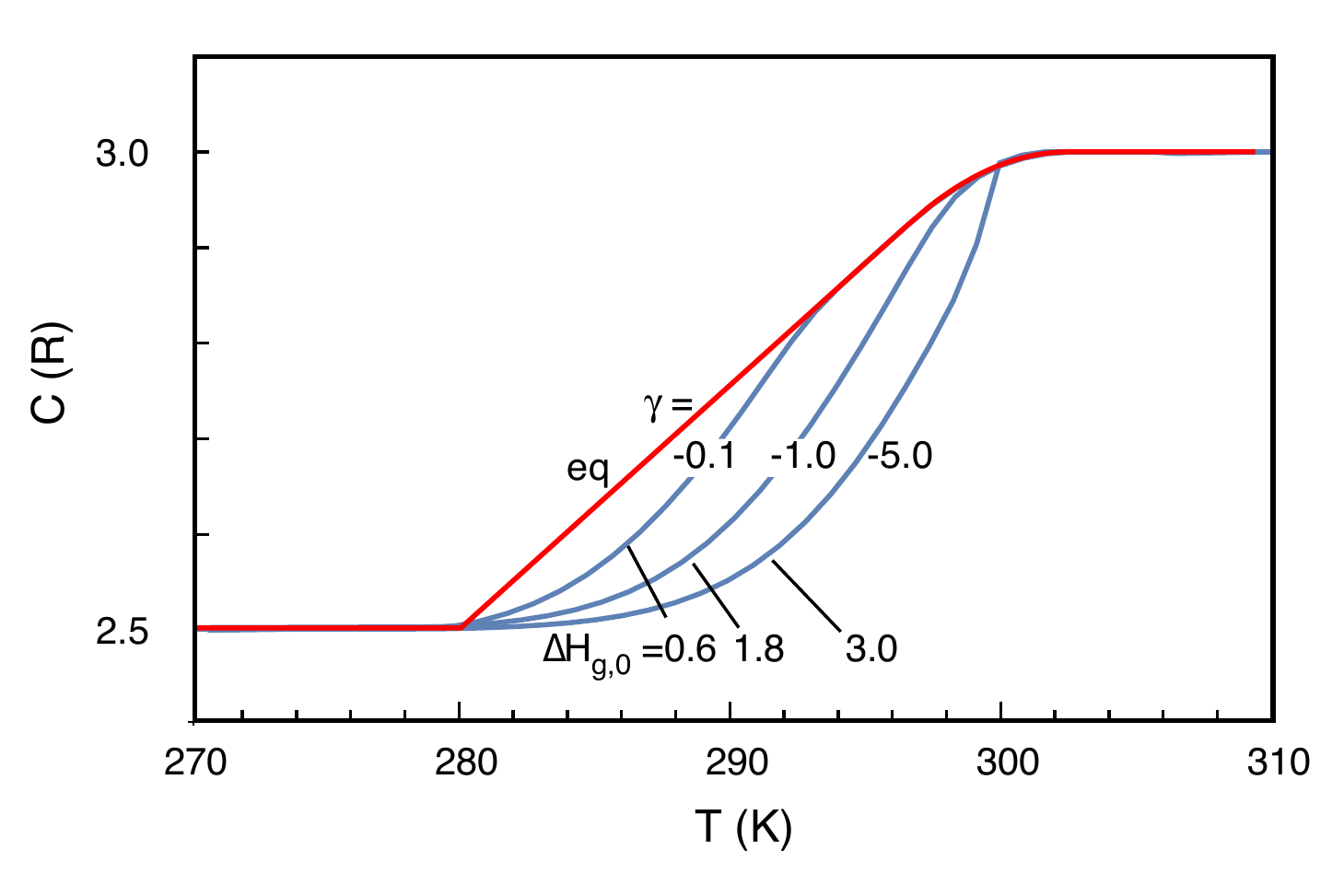} 
\includegraphics[width=80 mm, bb=0 0 420 280]{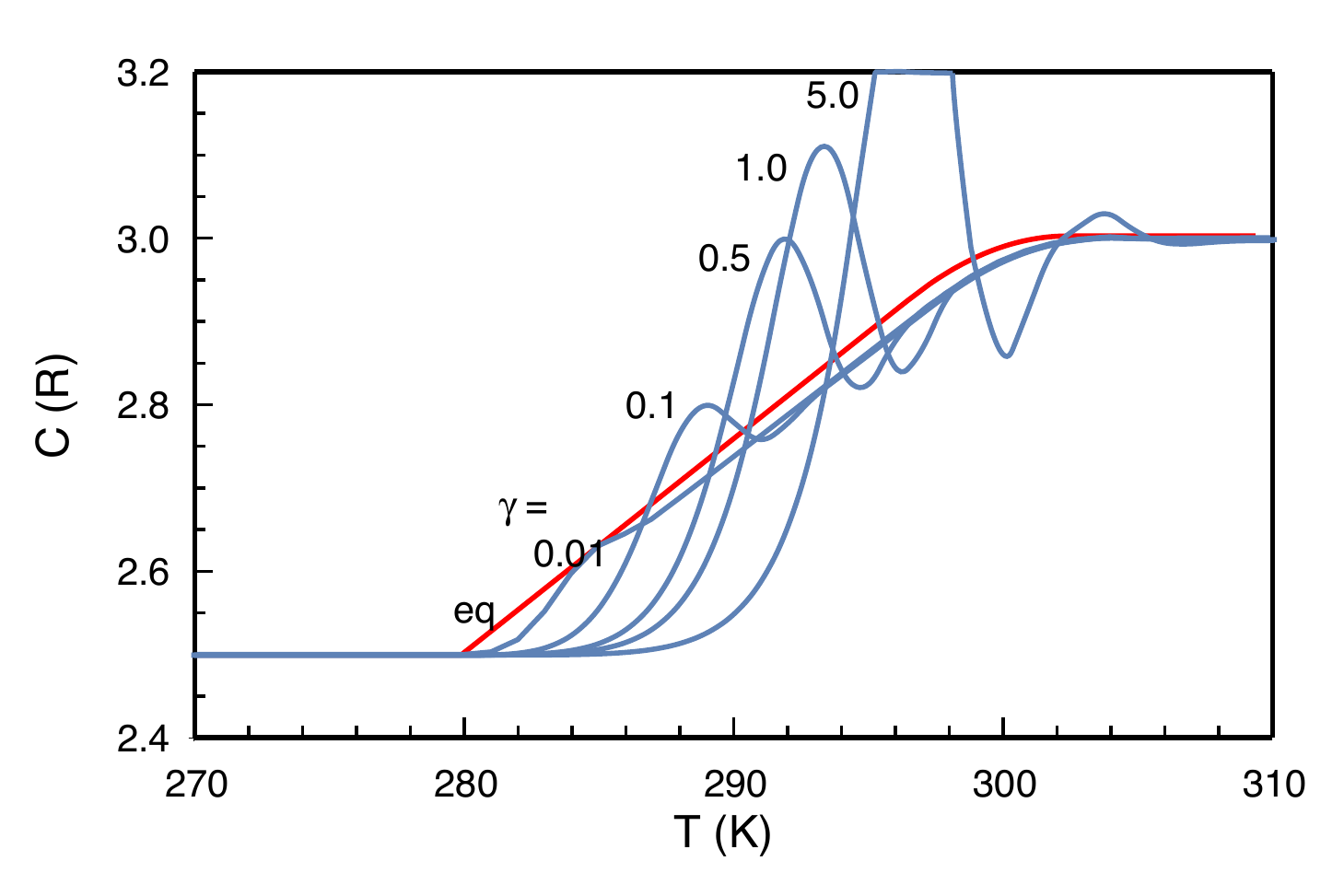} 
\caption{Specific-heat curves for cooling and heating processes. The cooling or heating rate $\gamma$ is presented in K/s. Red lines show the $C^{\rm (eq)}$-$T$ curve of the equilibrium state.}
\label{fig:rate-dep}
\end{figure}

Figure \ref{fig:rate-dep}(a) shows how the $C$-$T$ curve depends on the cooling rate $\gamma$ (${<}0$). As $| \gamma |$ increases, the specific heat $C$ drops faster, and the enthalpy of the glass $H_{g,0}$ at $T_{g,1}$ becomes larger. In this manner, the final state of the glass differs from sample to sample when different cooling rates are used. The deviation from the equilibrium value $\Delta H_{g,0}=H_{g,0}-H_{g,0}^{\rm (eq)}$ is called the residual enthalpy of the glass. Despite this, the specific heat of the glass is the same, because only $H_{t}$ contributes to specific heat $C$ at $T<T_{g,1}$.

Figure \ref{fig:rate-dep}(b) shows how the $C$-$T$ curve depends on the heating rate $\gamma$ (${>}0$) when starting from a sample in the lowest-energy state; that is, $\Delta H_{g,0}=0$. As the heating rate increases, the rise in specific heat $C$ shifts toward high temperatures. A hump appears for large $\gamma$, which is often found experimentally. This is because the structural relaxation cannot follow high rates of change in temperature $T$. The closed relation (\ref{eq:circ-heat0}) raises the $C$ near $T_{g,2}$ to compensate for the delay in the onset $T_{g,1}$. This creates a hump near $T_{g,2}$ when $\gamma$ is large.
In most experiments, the hump appears near $T_{g,2}$. However, the hump is observed even far from $T_{g,2}$, where the heating rate is very small \cite{Hodge83}. The present simulation indicates that this situation in fact occurs.


\begin{figure}[htbp]
\centering
\includegraphics[width=80 mm, bb=0 0 460 280]{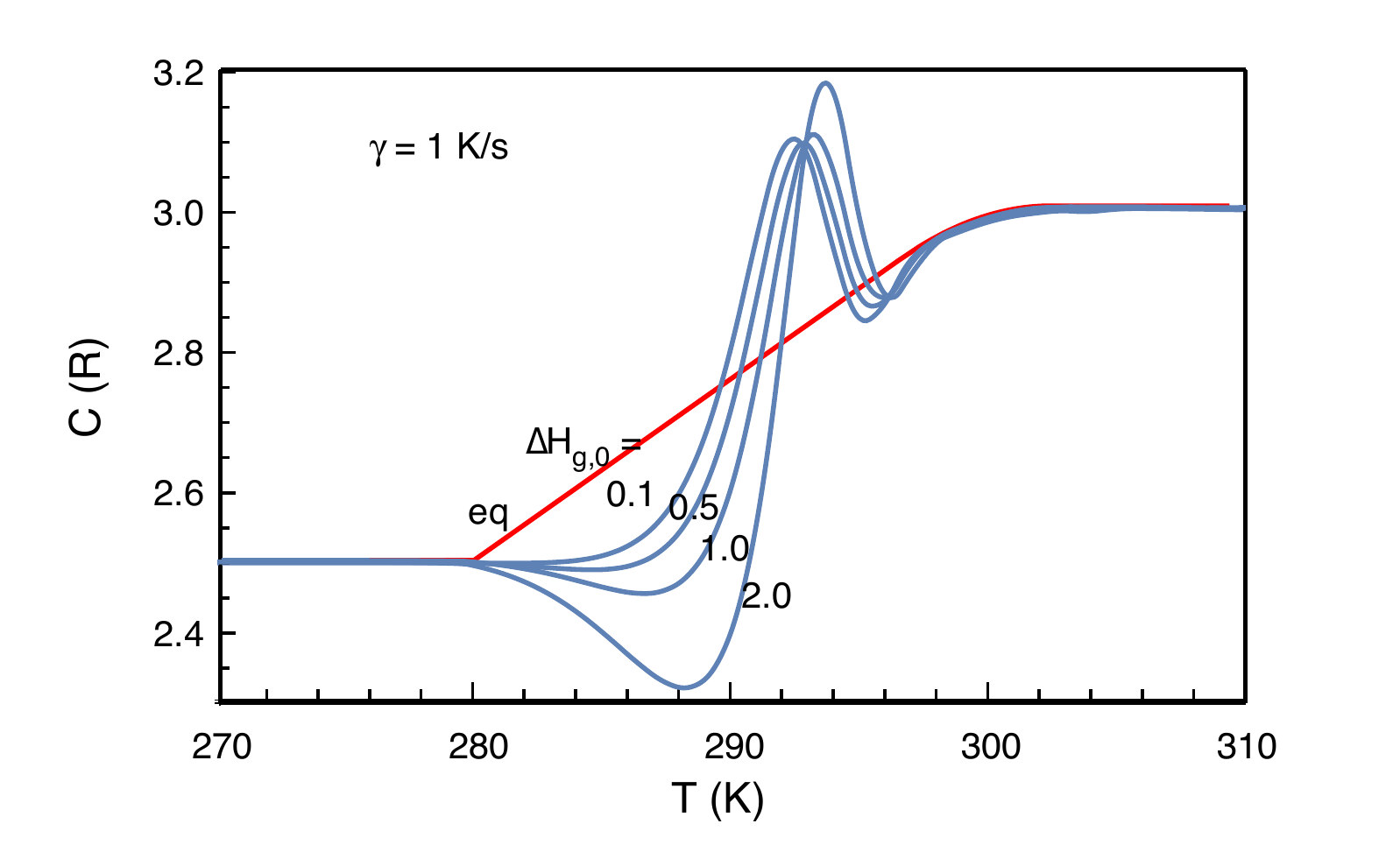} 
\caption{Effect of $\Delta H_{g,0}$ on the $C$-$T$ curve in the heating processes. The heating rate is fixed at $\gamma=1$~K/s. The red line shows the $C^{\rm (eq)}$-$T$ curve of the equilibrium state.}
\label{fig:differ-H0}
\end{figure}

Figure \ref{fig:differ-H0} shows how the $C$-$T$ curve depends on the residual enthalpy $H_{g,0}$. The sample is heated from the glass state at the rate $\gamma=1$~K/s. As the residual enthalpy $\Delta H_{g,0}$ increases, the increase in $C$ is delayed. This manifests a memory effect, which reflects the past history.
When $\Delta H_{g,0}$ is large, the specific heat $C$ initially decreases as $T$ increases. 
This behavior is observed experimentally for samples prepared at a very fast cooling rate \cite{Debolt76}, which implies a large residual enthalpy. 

In this manner, all the behaviors in the $C$-$T$ curve as $\gamma$ is changed are well-reproduced. One concern is the shape of the hump feature in the $C$-$T$ curve. When the sample is heating, experimental observations indicate that, subsequent to reaching the maximum in $C$, the $C$ converges monotonically to $C_{l}$ as $T$ increases further. However, the present simulations produce a somewhat oscillatory $C$-$T$ curve in that region. This issue is examined in the next section.

\subsubsection{Barrier spectroscopy}
We now do the calculation in the reverse direction to deduce the barrier height $E_b(H)$ from an observed $C$-$T$ curve. 
This approach may lead to a new method of spectroscopy to resolve how the barrier height $E_{b}$ depends on the enthalpy $H$, when the Arrhenius law breaks.
\begin{figure}[htbp]
\centering
\includegraphics[width=80 mm, bb=0 0 460 280]{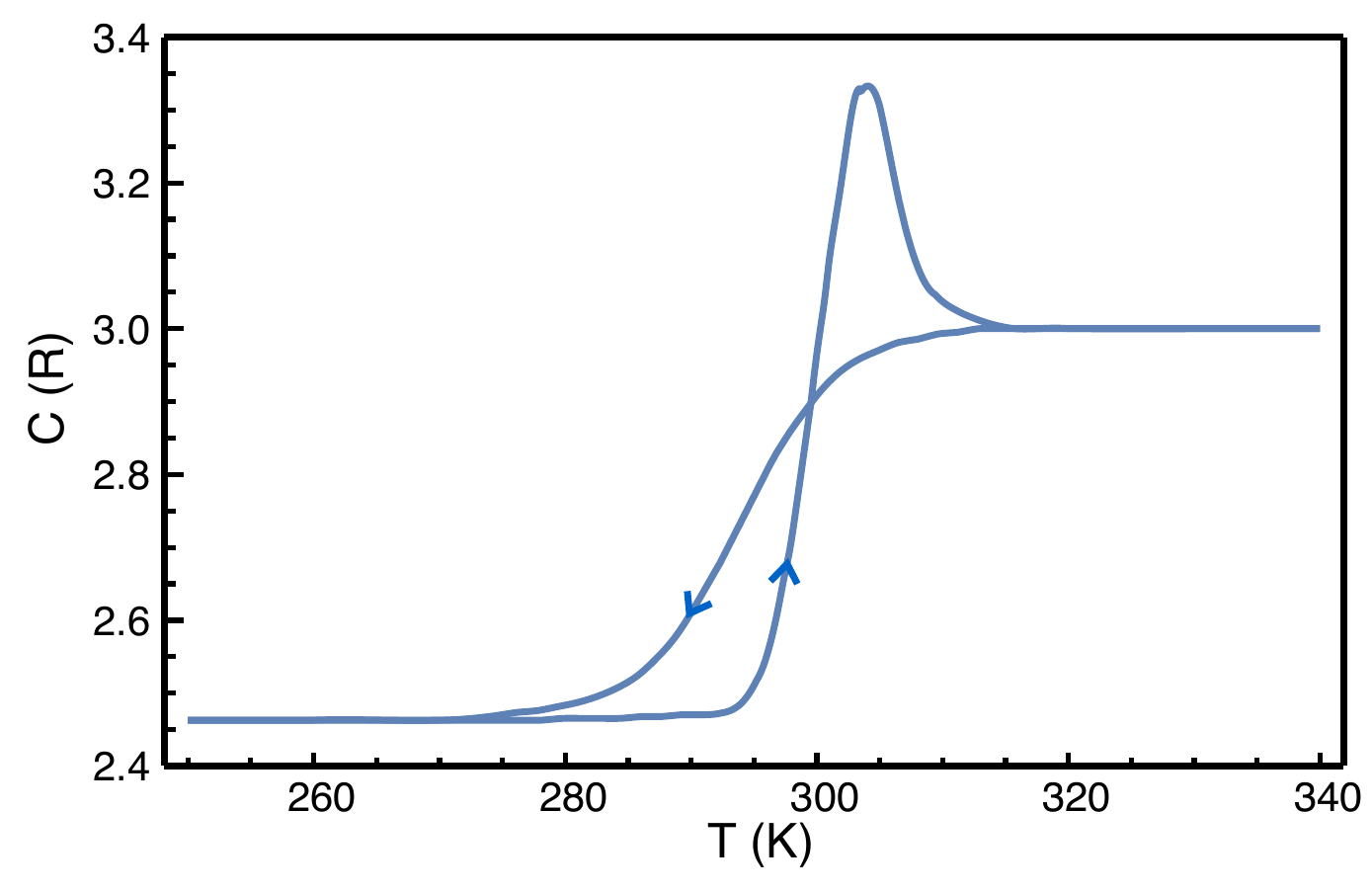} 
\includegraphics[width=80 mm, bb=0 0 460 280]{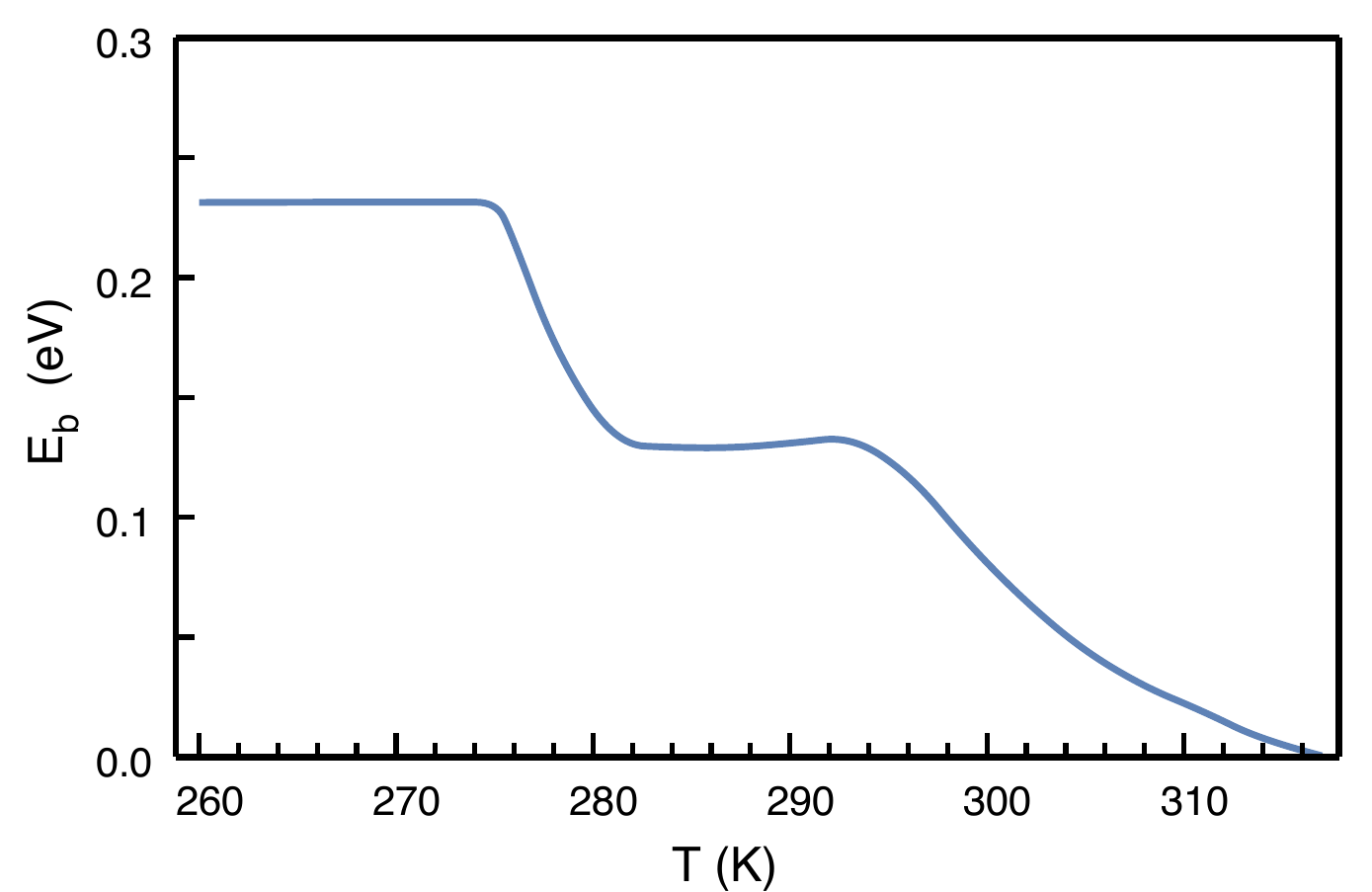} 
\caption{Barrier spectroscopy: (a) input data for $C$-$T$ curve, and (b) the resulting spectrum of barrier energy $E_{b}$.
} \label{fig:eb-spectra}
\end{figure}
Figure \ref{fig:eb-spectra}(a) shows the input data for the $C$-$T$ curve for a hypothetical glass. The specific heat $C$ is obtained by first cooling the sample as slow as possible from the liquid state to ensure that the specific heat $C$ is the equilibrium value $C^{\rm (eq)}$. After reaching the glass state, the sample is reheated at a finite rate $\gamma$. Although this curve does not represent any real data, it was prepared to meet the following conditions: (i) it exhibits a hump in the heating process, and (ii) it satisfies the condition (\ref{eq:circ-heat0}) over a single heating cycle, which ensures that the same sample is used.

Given the heating rate $\gamma=1$~K/s, $E_{b}$ is calculated and the result is plotted in Fig.~\ref{fig:eb-spectra}(b). An unexpected feature of this spectrum is the presence of a plateau in $E_{b}$ before the hump in $C$; otherwise $E_{b}$ decreases monotonically with $T$, as expected. Whether this is due to unrealistic input or to the inadequacy of the model to describe the relaxation process is unclear---remember that only a single relaxation time is assumed. 
Presently, we suppose that the input for the $C$-$T$ curve does not reflect the true $C$-$T$ curve of the glass state in the following sense: In the transition region, the glass substance is actually a mixture of the solid and liquid phases. The observed specific heat $C$ is thus the sum of the glass part and the liquid part. When $x$ mole fraction of the substance is in the liquid state, the observed specific heat $C$ becomes
\begin{equation}
C= (1-x) C_{g}+x C_{l}.
\label{eq:Cmix}
\end{equation}
Because the mole fraction $x$ of the liquid, which has no energy barrier $E_{b}$, increases as $T$ increases, the contribution of $C_{g}$ to the total $C$ decreases. Thus, applying Eq.~(\ref{eq:Eb-Hdep}) to the total $C$ leads to an overestimate of $E_{b}$ near $T_{g,2}$. Equation~(\ref{eq:Cmix}) must be taken into account to improve the simulation, which is left for a future study.

\section{Discussion of activation energy}
\label{sec:activation-energy}

Although a discussion of the details of the properties of glasses is not the main purpose of this paper, a few comments on the present results are warranted.
A significant question in the current study of glasses is the deviation of the activation energy from the Arrhenius law. Depending on the magnitude of the deviation, glasses may be classified as either strong or fragile \cite{Angell88,Angell95,Angell99}. 
Many authors have attempted to interpret the activation energy for the glass transition \cite{Adam65,Kovacs79,Scherer84,Hodge94,Nemilov-VitreousState, Johari00,Sastry01,Svoboda13}. 
In the present context, the change in the energy barrier $E_{b}$ as a function of $T$ is evident from the outset because $E_{b}$ is a function of atom position, $E_{b}=E_{b}(\{ \bar{\mathbf R}_{j} \})$. Although this is evident 
this is not reflected in the interpretation of the activation energy. (A $T$ dependence of the activation energy similar to the present model is found in the textbook of Nemilov \cite{Nemilov-VitreousState} (p.~161). Furthermore, this temperature dependence of the activation energy is consistent with the result of molecular-dynamic simulations by Debenedetti and Stillinger showing that the basins become deeper as the temperature $T$ decreases \cite{Debenedetti01}.)
For chemical reactions, the structures of the reactants and products do not change, and there is no reason that $E_{b}$ should change.
The question, therefore, is why $E_{b}$ appears not to deviate drastically from the Arrhenius law as is expected: for strong glasses the Arrhenius law holds well. The answer lies in the manner in which the Arrhenius plot is displayed. 
Equation~(\ref{eq:Eb-Hdep}) can be approximated as
\begin{equation}
E_{b}(T)=E_{b0}-b (T-T_{g,1}),
\label{eq:EbdepT} 
\end{equation}
where $b=E_{b0}/\Delta T_{g}$ is a constant. This approximation shows that $E_{b}$ varies significantly from $E_{b0}$ to zero over a narrow range of temperature $T$. However, in the Arrhenius plot, the term linear in $T$ is canceled by the linear term in the denominator in the exponent of Eq.~(\ref{eq:relax-tau}). Thus, Eq.~(\ref{eq:relax-tau}) becomes
\begin{equation}
\frac{1}{\tau} = A e^{b} \exp \left( -\frac{E_{b}^{\ast}}{k_{\rm B}T} \right),
\label{eq:relax-tau1}
\end{equation}
where $E_{b}^{\ast} = E_{b} + b T_{g,1}$.
This approximation shows why the linear term disappears from the Arrhenius analysis.

Furthermore, Eq.~(\ref{eq:EbdepT}) explains the large discrepancies in the related quantities.
First, the quantity obtained from the Arrhenius plot is $E_{b}^{\ast}$, not $E_{b}$. In the literature, the activation energy for the glass transitions of organic glasses are reported to be over 1 eV \cite{Hodge83,Hodge94}, and activation energies greater than 10 eV are not rare. Although the definition of activation energy slightly differs from method to method, these large values are surprising, as noted by Hodge \cite{Hodge91}. The energy barrier for the diffusion in metals is at most a few eV \cite{Mehrer07}. The migration enthalpy of a vacancy in silicon is about 0.1 eV \cite{Fahey89}. Based on the empirical rule for the activation enthalpy of diffusion, $E_{b}$ is given by
\begin{equation}
E_{b} = K_{1} T_{m},
\label{eq:eb-tm}
\end{equation}
where $K_{1}=1.5 \times 10^{-3} \ {\rm eV \, mol^{-1}\, K^{-1}}$ and $T_{m}$ is in K (\cite{Mehrer07}, p.~144). This gives a value $E_{b} = 0.6 \ {\rm eV/mol}$ for a material having the melting temperature $T_{m}=300$~K.
Although relaxation is a different phenomenon from diffusion, it is unrealistic that soft organic glasses, with $T_{m}$ often less than room temperature, have energy barriers larger than those of hard solids, even before complete solidification. In contrast, in Sec.~\ref{sec:heating}, $E_{b} =0.2$~eV was obtained, which indeed falls in a reasonable range in accordance with the empirical equation (\ref{eq:eb-tm}). The reason for this large discrepancy is that the quantity obtained by the Arrhenius plot is $E_{b}^{\ast}$, which is $E_{b}$ enlarged by a factor $b$. The coefficient $b=E_{b0}/\Delta T_{g}$ is very large because of the small value $\Delta T_{g}$. Thus, the term $b T_{g,1}$ becomes 2.8 eV for $T_{g,1}=280$~K, which exceeds the original value $E_{b0}$ by one order of magnitude.

Second, although the linear term cancels out, the effect of the coefficient $b$ remains in the pre-exponential factor $A^{\ast}=A e^{b}$. Even in the standard case for the Arrhenius analysis, it is not rare to find a quantitative discrepancy in the pre-exponential factor. In spite of this, a reasonable interpretation is possible. For example, for impurity diffusion, the pre-exponential factor $D_{0}$ in the diffusion constant $D=D_{0} \exp(-E_{b}/k_{\rm B}T)$ is related to the jumping frequency and jumping distance. This gives an order-of-magnitude agreement. 
Conversely, for the glass transitions, analyses \cite{Hodge83,Hodge94} show that $A$ is on the order of $10^{100} \text{ s}^{-1}$. No models can explain this extremely large value for $A$ by combining the elemental frequencies; for example, the typical phonon frequency is only $10^{14} \text{ s}^{-1}$. 
However, we have already seen that $b$ is very large, of the order of $10^{2}$. This factor is multiplied by $A$, giving a total pre-exponential factor $A^{\ast} \approx e^{100}$ s$^{-1}$. This is why quite large pre-exponential factors appear in the Arrhenius analyses.

Higher-order terms in Eq.~(\ref{eq:EbdepT}) makes the Arrhenius plot nonlinear, as found in the Angell plot \cite{Angell88,Angell95,Angell99}. The present theory may provide fresh insight into the Angell plot, which will be investigated in a future study.


\section{Conclusions} 
By defining a rigorous notion of equilibrium, the equilibrium positions of all constituent atoms, together with the internal energy, are shown to form a complete set of TCs for a given solid. By using a complete set of TCs, all the thermodynamic properties of a solid can be described by the present values of the TCs, irrespective of the past history of heat treatment.
Glasses are no exceptions. The only conclusion compatible with the thermodynamic principles is that the glass state is an equilibrium state. The fact that the properties of a glass depend on the previous states must be restated because, in fact, they depend only on the present atom positions $\{ \bar{\mathbf R}_{j} \}$, which are in fact the consequence of the past states.
Many of the internal variables or order parameters used in the literature can be qualified as TCs of glasses, if suitably treated.

Furthermore, under the approximation of the second kind, even a nonequilibrium phenomenon such as the glass transition can be described almost by the TCs only. This has been demonstrated through the analysis of a $C$-$T$ curve in the transition region. The only dynamic parameter involved in the model is the relaxation time $\tau_{s}$ for the structural relaxation. However, $\tau_{s}$ is expressed as a function of the present positions: $\tau_{s}(\{ \bar{\mathbf R}_{j} \})$. Therefore, no information of the previous states is needed. A main outcome of the proposed approach is the interpretation of the activation energy for the structural relaxation. The energy barrier for the structural relaxation depends strongly on the structure [$E_{b}(\{ \bar{\mathbf R}_{j} \})$]. Thus, the problem of the deviation from the Arrhenius law does not arise. The numerical estimates of $E_{b}$ resolve the unrealistic values previously reported.

A significant advantage of the present theory is that it makes no use of any hypothesis of what happens on long timescales, and hypothetical ideas, such as effective temperature, are not required. All quantities involved in the theory are accessible by experiment, and thus are verifiable in principle.


\begin{acknowledgments}
The following names are listed to thank for useful discussions with them: S. V. Nemilov (Saint-Petersburg National Research Univ. of Information Technologies), A. Takada (Asahi Glass Co. Ltd.), R. Conradt (RWTH Aachen Univ.), O. Yamamuro (Univ. Tokyo), and Y. Kajiwara (Hiroshima Univ). 
The author also thanks Enago (www.enago.jp) for the English language review.
This work was supported by the Research Program of ``Five-star Alliance" in ``NJRC Mater.~\& Dev."
\end{acknowledgments}

\appendix*
\section{ Free energy of transition state of glass} 

\label{sec:append-A}

\begin{figure}[htbp]
\centering
\includegraphics[width=70 mm, bb=0 0 360 250]{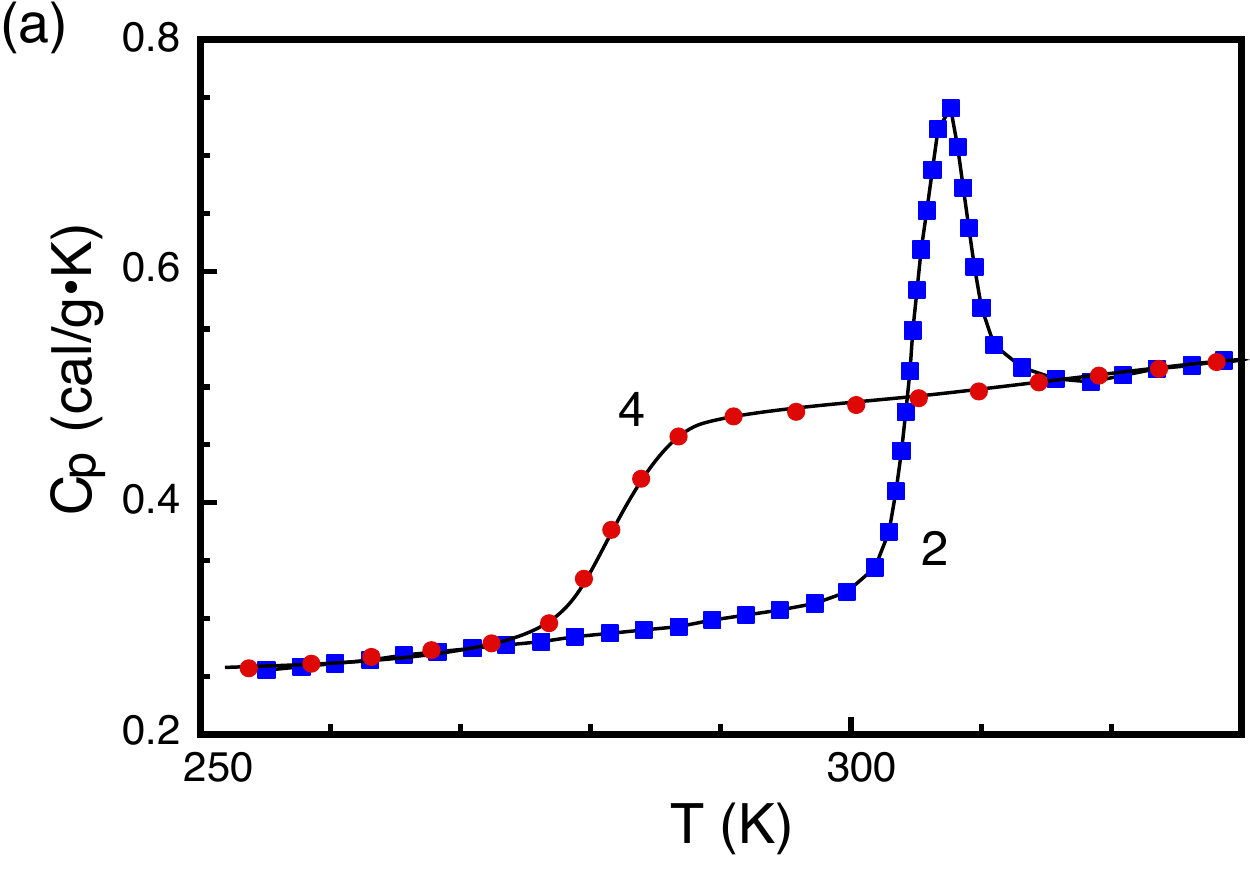} 
\hspace{5mm}
\includegraphics[width=68 mm, bb=0 0 350 250]{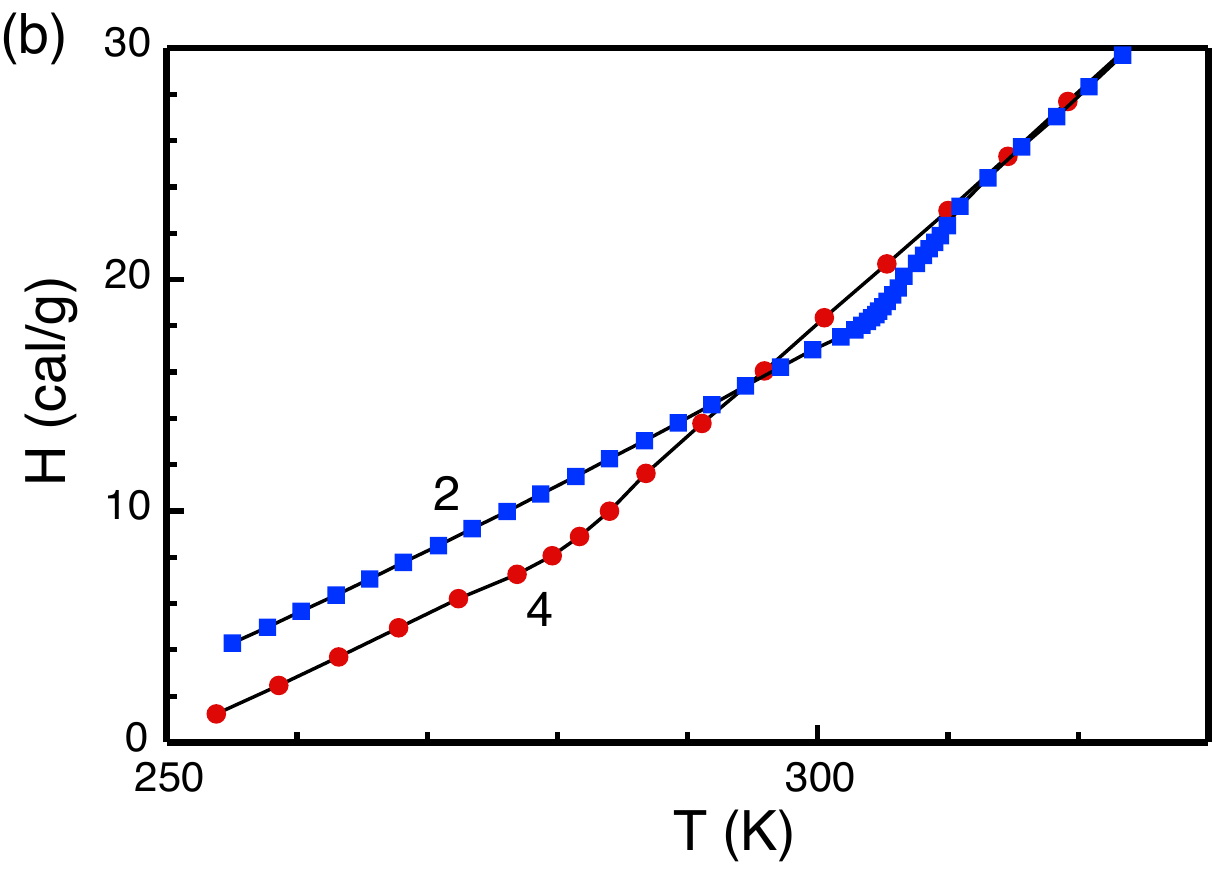} 

\vspace{5mm}
\includegraphics[width=70 mm, bb=0 0 360 250]{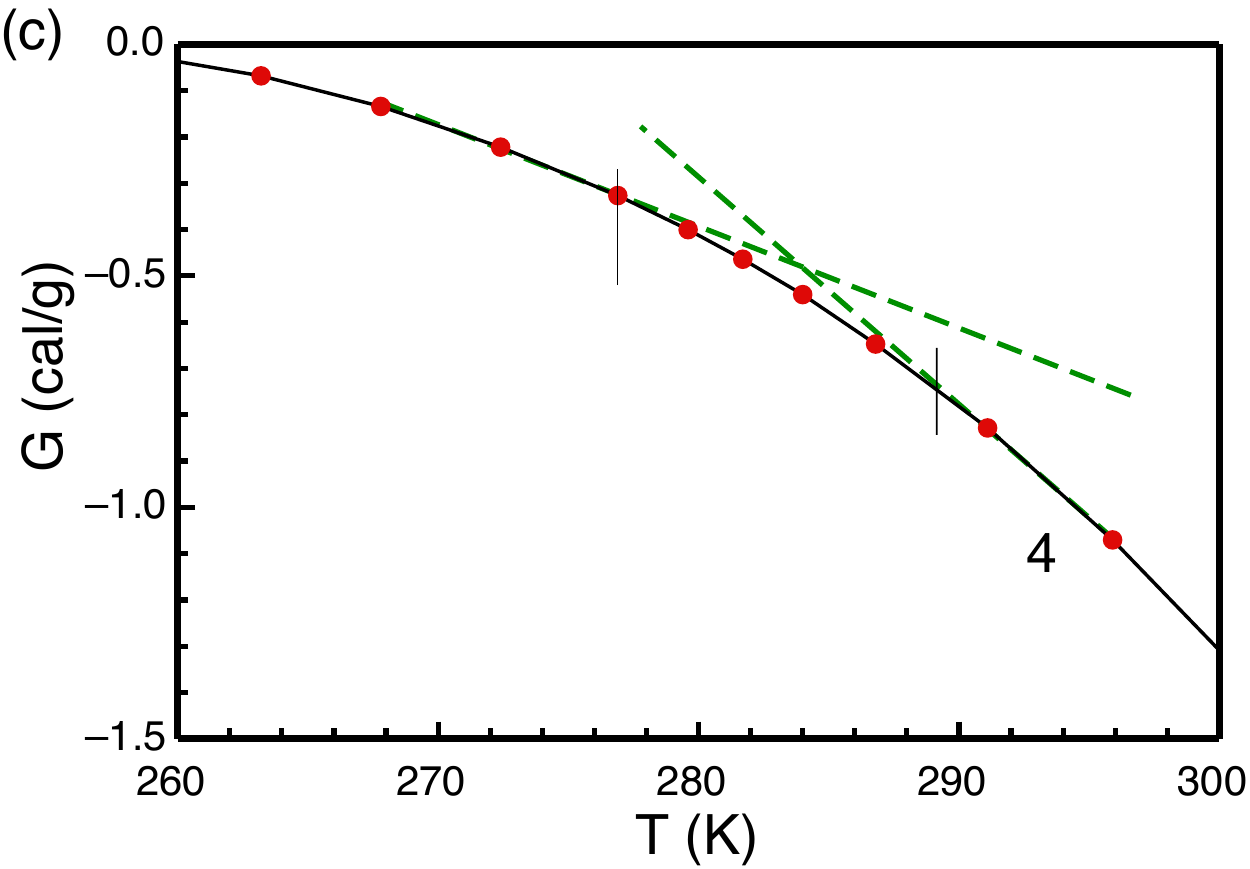} 
\hspace{5mm}
\includegraphics[width=70 mm, bb=0 0 360 250]{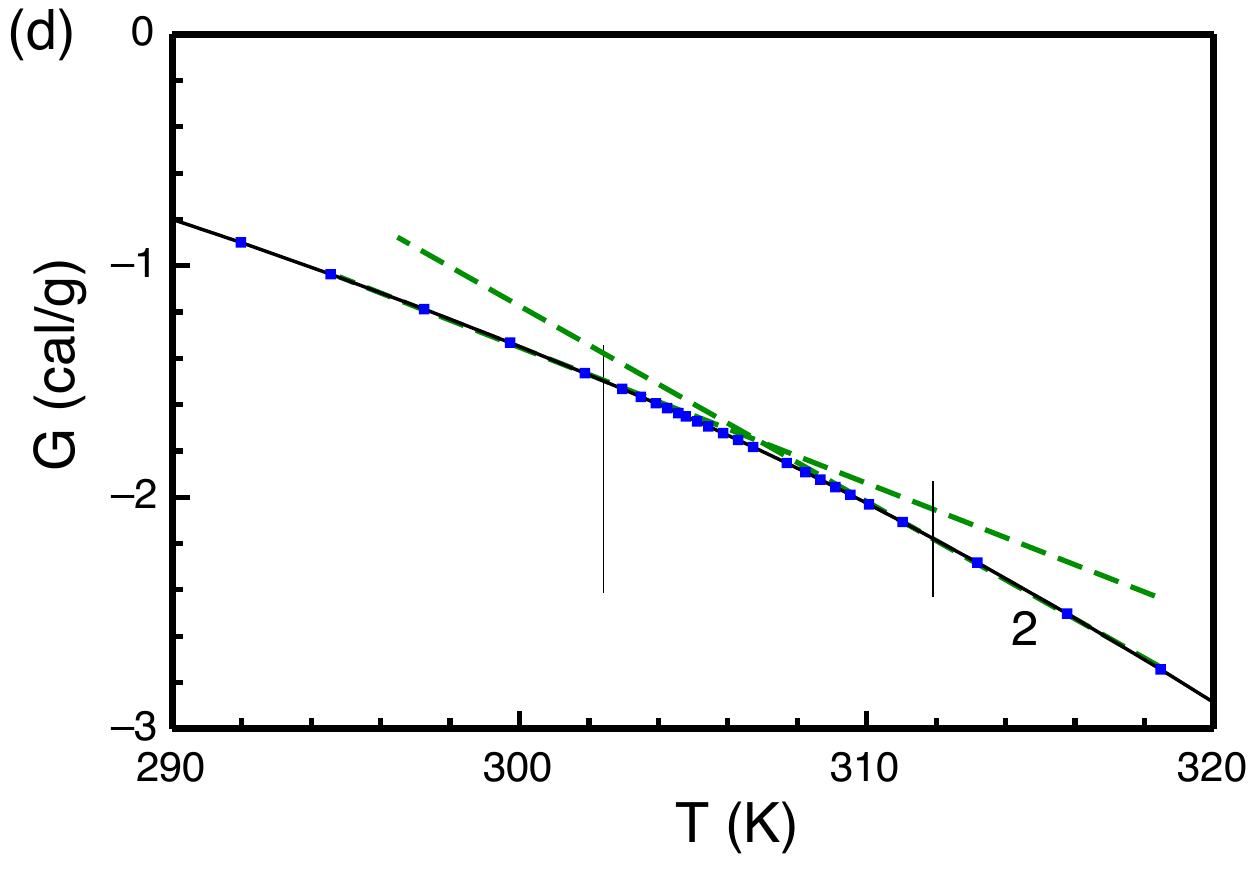} 
\caption{Isobaric specific heat $C_{p}$ of glucose. (a) Two data that have different thermal histories are digitized from Refs.~\cite{Parks28}(4, red) and \cite{Parks34} (2, blue). The specific-heat data were obtained by heating processes.
From the data $C_{p}$, the (b) enthalpy $H$, and (c), (d) Gibbs free energy $G$ are calculated. Note that different scales are used in panels (c) and (d) to plot $G$ in order to resolve the curvature in the transition regions (vertical bars). Extrapolations are indicated by green dashed lines. The origins of $H$, $S$, and $G$ are those values of data 4 at $T=250$ K.
} \label{fig:Cp-glucose}
\end{figure}

This appendix illustrates that the extrapolation of the enthalpy $H$ of a supercooled-liquid state leads to an incorrect conclusion. Consider Fig. \ref{fig:Cp-glucose}(a), which shows a $C$-$T$ curve of glucose.

Experimental data of the isobaric specific heat $C_{p}$ of glucose were taken from Refs.~\cite{Parks28} (4, red) and \cite{Parks34} (2, blue). The figures for $C_{p}$ were obtained by digitizing the curves from these papers. The enthalpy $H$ and Gibbs free energy $G$ were then obtained by integrating $C_{p}$ as $H(T)-H(T_{0})=\int_{T_{0}}^{T} C_{p}dT$, $S(T)-S(T_{0})=\int_{T_{0}}^{T} C_{p}d \ln T$, and similarly for $G(T)$. The reference temperature $T_{0}$ is taken to be a high temperature at which the glucose is in the liquid phase, which ensures that the thermodynamic functions of different samples are evaluated with the common origins. The original specific-heat measurements were done by heating. The heating rates $\gamma$ are 3$\,^{\circ}$C/h for data set 4 and 10$\,^{\circ}$C/h for data set 2. The two data sets for $C_{p}$ are scaled to match the value of the liquid state. Although the cooling rates for preparing the glass state of glucose are not explicitly given, these must be different, because the integration of $C_{p}$ over the transition region,  Eq.~(\ref{eq:circ-heat0}), does not give zero: there is a difference $\Delta H=2.8$ cal/g at $T=250$ K. Because data set 2 was obtained by using a faster heating rate, the hump near $T_{g,2}$ is reasonable.

As shown in both Figs.~\ref{fig:Cp-glucose}(c) and \ref{fig:Cp-glucose}(d), the free energy $G$ of the transition state is less than the extrapolated free energy of the supercooled liquid. This is a consequence of the positive-definite property of specific heat. This is a completely general property of stable material. Of course, it is also true that, at $T<T_{g,1}$, the free energy $G$ of the glass state is less than that of both the transition and supercooled-liquid states. The metastability of the glass state has never been proven.


\begin{thebibliography}{103}
\expandafter\ifx\csname natexlab\endcsname\relax\def\natexlab#1{#1}\fi
\expandafter\ifx\csname bibnamefont\endcsname\relax
  \def\bibnamefont#1{#1}\fi
\expandafter\ifx\csname bibfnamefont\endcsname\relax
  \def\bibfnamefont#1{#1}\fi
\expandafter\ifx\csname citenamefont\endcsname\relax
  \def\citenamefont#1{#1}\fi
\expandafter\ifx\csname url\endcsname\relax
  \def\url#1{\texttt{#1}}\fi
\expandafter\ifx\csname urlprefix\endcsname\relax\def\urlprefix{URL }\fi
\providecommand{\bibinfo}[2]{#2}
\providecommand{\eprint}[2][]{\url{#2}}

\bibitem[{\citenamefont{Bridgman}(1961)}]{Bridgman41}
\bibinfo{author}{\bibfnamefont{P.~W.} \bibnamefont{Bridgman}},
  \emph{\bibinfo{title}{The Nature of Thermodynamics}}
  (\bibinfo{publisher}{Harper \& Brothers}, \bibinfo{address}{New York},
  \bibinfo{year}{1961}).

\bibitem[{\citenamefont{Bridgman}(1950)}]{Bridgman50}
\bibinfo{author}{\bibfnamefont{P.~W.} \bibnamefont{Bridgman}},
  \bibinfo{journal}{Rev. Mod. Phys.} \textbf{\bibinfo{volume}{23}},
  \bibinfo{pages}{56} (\bibinfo{year}{1950}).

\bibitem[{com()}]{comment-equilibrium}
\bibinfo{note}{Even for a reading scientist, giving a consistent definition of
  equilibrium is difficult. Tisza once proposed the unorthodox view that the
  third law can be used for the criterion of equilibrium: see, L. Tisza, {\it
  Generalized Thermodynamics} (MIT Press, Cambridge, 1966).}

\bibitem[{\citenamefont{Gyftopoulos and Beretta}(2005)}]{Gyftopoulos}
\bibinfo{author}{\bibfnamefont{E.~P.} \bibnamefont{Gyftopoulos}}
  \bibnamefont{and} \bibinfo{author}{\bibfnamefont{G.~P.}
  \bibnamefont{Beretta}}, \emph{\bibinfo{title}{Thermodynamics - Foundations
  and Applications}} (\bibinfo{publisher}{Dover Pub.}, \bibinfo{address}{New
  York}, \bibinfo{year}{2005}).

\bibitem[{Sta()}]{StateVariable}
\bibinfo{note}{K. Shirai, cond-mat.stat-mech/1812.08977}.

\bibitem[{\citenamefont{Zemansky and Dittman}(1997)}]{Zemansky}
\bibinfo{author}{\bibfnamefont{M.}~\bibnamefont{Zemansky}} \bibnamefont{and}
  \bibinfo{author}{\bibfnamefont{R.}~\bibnamefont{Dittman}},
  \emph{\bibinfo{title}{Heat and Thermodynamics}}
  (\bibinfo{publisher}{McGraw-Hill}, \bibinfo{address}{New York},
  \bibinfo{year}{1997}), \bibinfo{edition}{7th} ed.

\bibitem[{\citenamefont{Callen}(1985)}]{Callen}
\bibinfo{author}{\bibfnamefont{H.}~\bibnamefont{Callen}},
  \emph{\bibinfo{title}{Thermodynamics and an Introduction to
  Thermostatistics}} (\bibinfo{publisher}{Wiley}, \bibinfo{address}{New York},
  \bibinfo{year}{1985}), \bibinfo{edition}{2nd} ed.

\bibitem[{\citenamefont{Simon}(1930)}]{Simon30}
\bibinfo{author}{\bibfnamefont{F.~E.} \bibnamefont{Simon}},
  \bibinfo{journal}{Ergebn. exalct. Naturwiss} \textbf{\bibinfo{volume}{9}},
  \bibinfo{pages}{244} (\bibinfo{year}{1930}).

\bibitem[{\citenamefont{Fowler and Guggenheim}(1952)}]{Fowler-Guggenheim}
\bibinfo{author}{\bibfnamefont{R.}~\bibnamefont{Fowler}} \bibnamefont{and}
  \bibinfo{author}{\bibfnamefont{E.~A.} \bibnamefont{Guggenheim}},
  \emph{\bibinfo{title}{Statistical Thermodynamics}}
  (\bibinfo{publisher}{Cambridge}, \bibinfo{address}{London},
  \bibinfo{year}{1952}), \bibinfo{edition}{3rd} ed.

\bibitem[{\citenamefont{Davies and Jones}(1953{\natexlab{a}})}]{Davies53a}
\bibinfo{author}{\bibfnamefont{R.~O.} \bibnamefont{Davies}} \bibnamefont{and}
  \bibinfo{author}{\bibfnamefont{G.~O.} \bibnamefont{Jones}},
  \bibinfo{journal}{Adv. Phys.} \textbf{\bibinfo{volume}{2}},
  \bibinfo{pages}{370} (\bibinfo{year}{1953}{\natexlab{a}}).

\bibitem[{\citenamefont{Jackle}(1986)}]{Jackle86}
\bibinfo{author}{\bibfnamefont{J.}~\bibnamefont{Jackle}},
  \bibinfo{journal}{Rep. Prog. Phys.} \textbf{\bibinfo{volume}{49}},
  \bibinfo{pages}{171} (\bibinfo{year}{1986}).

\bibitem[{\citenamefont{Angell et~al.}(1999)\citenamefont{Angell, Richards, and
  Velikov}}]{Angell99}
\bibinfo{author}{\bibfnamefont{C.~A.} \bibnamefont{Angell}},
  \bibinfo{author}{\bibfnamefont{B.~E.} \bibnamefont{Richards}},
  \bibnamefont{and} \bibinfo{author}{\bibfnamefont{V.}~\bibnamefont{Velikov}},
  \bibinfo{journal}{J. Phys.: Condens. Matter} \textbf{\bibinfo{volume}{11}},
  \bibinfo{pages}{A75} (\bibinfo{year}{1999}).

\bibitem[{\citenamefont{Rao}(2002)}]{Rao02}
\bibinfo{author}{\bibfnamefont{K.~J.} \bibnamefont{Rao}},
  \emph{\bibinfo{title}{Structural chemistry of glasses}}
  (\bibinfo{publisher}{Elsevier}, \bibinfo{address}{Amsterdam},
  \bibinfo{year}{2002}).

\bibitem[{\citenamefont{Mysen and Richet}(2005)}]{Mysen-Richet05}
\bibinfo{author}{\bibfnamefont{B.}~\bibnamefont{Mysen}} \bibnamefont{and}
  \bibinfo{author}{\bibfnamefont{P.}~\bibnamefont{Richet}},
  \emph{\bibinfo{title}{Silicate Glasses and Melts: Properties and Structure}}
  (\bibinfo{publisher}{Elsevier}, \bibinfo{address}{Amsterdam},
  \bibinfo{year}{2005}).

\bibitem[{\citenamefont{Berthier and Biroli}(2011)}]{Berthier11}
\bibinfo{author}{\bibfnamefont{L.}~\bibnamefont{Berthier}} \bibnamefont{and}
  \bibinfo{author}{\bibfnamefont{G.}~\bibnamefont{Biroli}},
  \bibinfo{journal}{Rev. Mod. Phys.} \textbf{\bibinfo{volume}{83}},
  \bibinfo{pages}{587} (\bibinfo{year}{2011}).

\bibitem[{\citenamefont{Biroli and Garrahan}(2013)}]{Biroli13}
\bibinfo{author}{\bibfnamefont{G.}~\bibnamefont{Biroli}} \bibnamefont{and}
  \bibinfo{author}{\bibfnamefont{P.}~\bibnamefont{Garrahan}},
  \bibinfo{journal}{J. Chem. Phys.} \textbf{\bibinfo{volume}{138}},
  \bibinfo{pages}{12A301} (\bibinfo{year}{2013}).

\bibitem[{\citenamefont{Berthier and Ediger}(2016)}]{Berthier16}
\bibinfo{author}{\bibfnamefont{L.}~\bibnamefont{Berthier}} \bibnamefont{and}
  \bibinfo{author}{\bibfnamefont{M.~D.} \bibnamefont{Ediger}},
  \bibinfo{journal}{Phys. Today} \textbf{\bibinfo{volume}{69}},
  \bibinfo{pages}{(1) 40} (\bibinfo{year}{2016}).

\bibitem[{\citenamefont{Kauzmann}(1948)}]{Kauzmann48}
\bibinfo{author}{\bibfnamefont{W.}~\bibnamefont{Kauzmann}},
  \bibinfo{journal}{Chem. Rev.} \textbf{\bibinfo{volume}{43}},
  \bibinfo{pages}{219} (\bibinfo{year}{1948}).

\bibitem[{\citenamefont{Zhang et~al.}(2019)\citenamefont{Zhang, Mazzarello, and
  Ma}}]{MRS-PCM19}
\bibinfo{author}{\bibfnamefont{W.}~\bibnamefont{Zhang}},
  \bibinfo{author}{\bibfnamefont{R.}~\bibnamefont{Mazzarello}},
  \bibnamefont{and} \bibinfo{author}{\bibfnamefont{E.}~\bibnamefont{Ma}},
  \bibinfo{journal}{MRS Bull.} \textbf{\bibinfo{volume}{44}},
  \bibinfo{pages}{686} (\bibinfo{year}{2019}), \bibinfo{note}{the papers in this
  special issue}.

\bibitem[{not()}]{note1}
\bibinfo{note}{The author is unable to follow diverse developments in these
  fields, so that only a few are cited here: 
  H. Mamiya and S. Nimori, J. Appl. Phys. {\bf 111}, 07E147 (2012);
  A. Samarakoon, T. J. Sato, T. Chen, G.-W. Chern, J. Yang, I. Klich, R. Sinchair, H. Zhou,
and S. Lee, Proc. Nat. Acad. Sci. {\bf 113}, 11806 (2016).
  }.

\bibitem[{\citenamefont{Gibbs}(1906)}]{Gibbs}
\bibinfo{author}{\bibfnamefont{J.~W.} \bibnamefont{Gibbs}},
  \emph{\bibinfo{title}{Scientific Papers}}, vol. \bibinfo{volume}{I:
  Thermodynamics} (\bibinfo{publisher}{Longmans, Green and Co.},
  \bibinfo{address}{New York}, \bibinfo{year}{1906}).

\bibitem[{\citenamefont{Reiss}(1996)}]{Reiss}
\bibinfo{author}{\bibfnamefont{H.}~\bibnamefont{Reiss}},
  \emph{\bibinfo{title}{Methods of Thermodynamics}} (\bibinfo{publisher}{Dover
  Pub.}, \bibinfo{address}{New York}, \bibinfo{year}{1996}).

\bibitem[{\citenamefont{Ben-Naim}(2008)}]{Ben-Naim}
\bibinfo{author}{\bibfnamefont{A.}~\bibnamefont{Ben-Naim}},
  \emph{\bibinfo{title}{Entropy Demystified}} (\bibinfo{publisher}{World
  Scientific}, \bibinfo{address}{Singapore}, \bibinfo{year}{2008}),
  \bibinfo{edition}{expanded} ed.

\bibitem[{\citenamefont{ter Haar}(1966)}]{Haar-Thermostat}
\bibinfo{author}{\bibfnamefont{D.}~\bibnamefont{ter Haar}},
  \emph{\bibinfo{title}{Elements of Thermostatistics}}
  (\bibinfo{publisher}{Holt, Rinehart and Winston}, \bibinfo{address}{New
  York}, \bibinfo{year}{1966}), \bibinfo{edition}{2nd} ed.

\bibitem[{\citenamefont{Rosenkrantz}(1983)}]{Rosenkrantz83}
\bibinfo{editor}{\bibfnamefont{R.~D.} \bibnamefont{Rosenkrantz}}, ed.,
  \emph{\bibinfo{title}{E. T. Jaynes: Papers on Probability, Statistics and
  Statistical Physics}} (\bibinfo{publisher}{Reidel Pub.},
  \bibinfo{address}{Dordrecht}, \bibinfo{year}{1983}).

\bibitem[{\citenamefont{Hatsopoulos and Keenan}(1965)}]{Hatsopoulos}
\bibinfo{author}{\bibfnamefont{G.~N.} \bibnamefont{Hatsopoulos}}
  \bibnamefont{and} \bibinfo{author}{\bibfnamefont{J.~H.}
  \bibnamefont{Keenan}}, \emph{\bibinfo{title}{Principles of General
  Thermodynamics}} (\bibinfo{publisher}{John Wiley \& Sons, Inc.},
  \bibinfo{address}{New York}, \bibinfo{year}{1965}).

\bibitem[{\citenamefont{Callaway and March}(1984)}]{Callaway84}
\bibinfo{author}{\bibfnamefont{J.}~\bibnamefont{Callaway}} \bibnamefont{and}
  \bibinfo{author}{\bibfnamefont{N.~H.} \bibnamefont{March}},
  \bibinfo{journal}{Solid State Physics, eds. H. Ehrenreich and D. Turnbull}
  \textbf{\bibinfo{volume}{38}}, \bibinfo{pages}{135} (\bibinfo{year}{1984}).

\bibitem[{\citenamefont{Parr and Yang}(1989)}]{Parr-Yang89}
\bibinfo{author}{\bibfnamefont{R.~G.} \bibnamefont{Parr}} \bibnamefont{and}
  \bibinfo{author}{\bibfnamefont{W.}~\bibnamefont{Yang}},
  \emph{\bibinfo{title}{Density-Functional Theory of Atoms and Molecules}}
  (\bibinfo{publisher}{Oxford}, \bibinfo{address}{Oxford},
  \bibinfo{year}{1989}).

\bibitem[{\citenamefont{Zangwill}(2015)}]{Zangwill15}
\bibinfo{author}{\bibfnamefont{A.}~\bibnamefont{Zangwill}},
  \bibinfo{journal}{Phys. Today} \textbf{\bibinfo{volume}{68}},
  \bibinfo{pages}{(7) 34} (\bibinfo{year}{2015}).

\bibitem[{\citenamefont{Born and Huang}(1969)}]{BornHuang}
\bibinfo{author}{\bibfnamefont{M.}~\bibnamefont{Born}} \bibnamefont{and}
  \bibinfo{author}{\bibfnamefont{K.}~\bibnamefont{Huang}},
  \emph{\bibinfo{title}{Dynamical Theory of Crystal Lattices}}
  (\bibinfo{publisher}{Clarendon}, \bibinfo{address}{Oxford},
  \bibinfo{year}{1969}).

\bibitem[{\citenamefont{Mandelbrot}(1964)}]{Mandelbrot64}
\bibinfo{author}{\bibfnamefont{B.}~\bibnamefont{Mandelbrot}},
  \bibinfo{journal}{J. Math. Phys.} \textbf{\bibinfo{volume}{5}},
  \bibinfo{pages}{164} (\bibinfo{year}{1964}).

\bibitem[{\citenamefont{Tisza and Quay}(1963)}]{Tisza63}
\bibinfo{author}{\bibfnamefont{L.}~\bibnamefont{Tisza}} \bibnamefont{and}
  \bibinfo{author}{\bibfnamefont{P.~M.} \bibnamefont{Quay}},
  \bibinfo{journal}{Ann. Phys} \textbf{\bibinfo{volume}{25}},
  \bibinfo{pages}{48} (\bibinfo{year}{1963}).

\bibitem[{\citenamefont{den Broeck and Esposito}(2015)}]{Broeck15}
\bibinfo{author}{\bibfnamefont{C.~} \bibnamefont{Van den Broeck}}
  \bibnamefont{and} \bibinfo{author}{\bibfnamefont{M.}~\bibnamefont{Esposito}},
  \bibinfo{journal}{Physica A} \textbf{\bibinfo{volume}{418}},
  \bibinfo{pages}{6} (\bibinfo{year}{2015}).

\bibitem[{\citenamefont{Penrose}(1979)}]{Penrose79}
\bibinfo{author}{\bibfnamefont{O.}~\bibnamefont{Penrose}},
  \bibinfo{journal}{Rep. Prog. Phys.} \textbf{\bibinfo{volume}{42}},
  \bibinfo{pages}{129} (\bibinfo{year}{1979}).

\bibitem[{\citenamefont{Lebowitz}(1993)}]{Lebowitz93}
\bibinfo{author}{\bibfnamefont{J.~L.} \bibnamefont{Lebowitz}},
  \bibinfo{journal}{Phys. Today} \textbf{\bibinfo{volume}{46}},
  \bibinfo{pages}{(9) 32} (\bibinfo{year}{1993}).

\bibitem[{\citenamefont{Mackey}(1989)}]{Mackey89}
\bibinfo{author}{\bibfnamefont{M.~C.} \bibnamefont{Mackey}},
  \bibinfo{journal}{Rev. Mod. Phys.} \textbf{\bibinfo{volume}{61}},
  \bibinfo{pages}{981} (\bibinfo{year}{1989}).

\bibitem[{\citenamefont{Uffink}(2001)}]{Uffink01}
\bibinfo{author}{\bibfnamefont{J.}~\bibnamefont{Uffink}},
  \bibinfo{journal}{Stud. Hist. Phil. Mod. Phys.}
  \textbf{\bibinfo{volume}{32}}, \bibinfo{pages}{305} (\bibinfo{year}{2001}).

\bibitem[{\citenamefont{Jr.}(2008)}]{Grandy}
\bibinfo{author}{\bibfnamefont{W.~T.~Grandy} \bibnamefont{Jr.}},
  \emph{\bibinfo{title}{Entropy and the Time Evolution of Macroscopic Systems}}
  (\bibinfo{publisher}{Oxford}, \bibinfo{address}{Oxford},
  \bibinfo{year}{2008}).

\bibitem[{\citenamefont{Boksenbojm et~al.}(2011)\citenamefont{Boksenbojm,
  Wynants, and Jarzynski}}]{Boksenbojm11}
\bibinfo{author}{\bibfnamefont{E.}~\bibnamefont{Boksenbojm}},
  \bibinfo{author}{\bibfnamefont{B.}~\bibnamefont{Wynants}}, \bibnamefont{and}
  \bibinfo{author}{\bibfnamefont{C.}~\bibnamefont{Jarzynski}},
  \bibinfo{journal}{Annu. Rev. Condens. Matter Phys.}
  \textbf{\bibinfo{volume}{2}}, \bibinfo{pages}{329} (\bibinfo{year}{2011}).

\bibitem[{\citenamefont{Lieb and Yngvason}(2013)}]{Lieb13}
\bibinfo{author}{\bibfnamefont{E.~H.} \bibnamefont{Lieb}} \bibnamefont{and}
  \bibinfo{author}{\bibfnamefont{J.}~\bibnamefont{Yngvason}},
  \bibinfo{journal}{Proc. R. Soc. A} \textbf{\bibinfo{volume}{469}},
  \bibinfo{pages}{20130408} (\bibinfo{year}{2013}).

\bibitem[{\citenamefont{Swendsen}(2017)}]{Swendsen17}
\bibinfo{author}{\bibfnamefont{R.~H.} \bibnamefont{Swendsen}},
  \bibinfo{journal}{Physica A} \textbf{\bibinfo{volume}{467}},
  \bibinfo{pages}{67} (\bibinfo{year}{2017}).

\bibitem[{\citenamefont{Prigogine and Defay}(1954)}]{Prigogine54}
\bibinfo{author}{\bibfnamefont{I.}~\bibnamefont{Prigogine}} \bibnamefont{and}
  \bibinfo{author}{\bibfnamefont{R.}~\bibnamefont{Defay}},
  \emph{\bibinfo{title}{Chemical Thermodynamics}}
  (\bibinfo{publisher}{Longmans}, \bibinfo{address}{London},
  \bibinfo{year}{1954}).

\bibitem[{\citenamefont{Bouchbinder and Langer}(2009)}]{Bouchbinder09-1}
\bibinfo{author}{\bibfnamefont{E.}~\bibnamefont{Bouchbinder}} \bibnamefont{and}
  \bibinfo{author}{\bibfnamefont{J.~S.} \bibnamefont{Langer}},
  \bibinfo{journal}{Phys. Rev. E} \textbf{\bibinfo{volume}{80}},
  \bibinfo{pages}{031131} (\bibinfo{year}{2009}).

\bibitem[{\citenamefont{Gujrati}(2010)}]{Gujrati10}
\bibinfo{author}{\bibfnamefont{P.~D.} \bibnamefont{Gujrati}},
  \bibinfo{journal}{Phys. Rev. E} \textbf{\bibinfo{volume}{81}},
  \bibinfo{pages}{051130} (\bibinfo{year}{2010}).

\bibitem[{\citenamefont{Sciortino}(2005)}]{Sciortino05}
\bibinfo{author}{\bibfnamefont{F.}~\bibnamefont{Sciortino}},
  \bibinfo{journal}{J. Stat. Mech.} \textbf{\bibinfo{volume}{2005}},
  \bibinfo{pages}{P05015} (\bibinfo{year}{2005}).

\bibitem[{\citenamefont{Nieuwenhuizen}(1998)}]{Nieuwenhuizen98a}
\bibinfo{author}{\bibfnamefont{T.~M.} \bibnamefont{Nieuwenhuizen}},
  \bibinfo{journal}{Phys. Rev. Lett.} \textbf{\bibinfo{volume}{80}},
  \bibinfo{pages}{5580} (\bibinfo{year}{1998}).

\bibitem[{\citenamefont{Davies and Jones}(1953{\natexlab{b}})}]{Davies53}
\bibinfo{author}{\bibfnamefont{R.~O.} \bibnamefont{Davies}} \bibnamefont{and}
  \bibinfo{author}{\bibfnamefont{G.~O.} \bibnamefont{Jones}},
  \bibinfo{journal}{Proc. Roy. Soc. A} \textbf{\bibinfo{volume}{217}},
  \bibinfo{pages}{26} (\bibinfo{year}{1953}{\natexlab{b}}).

\bibitem[{\citenamefont{Stillinger}(1995)}]{Stillinger95}
\bibinfo{author}{\bibfnamefont{F.~H.} \bibnamefont{Stillinger}},
  \bibinfo{journal}{Science} \textbf{\bibinfo{volume}{267}},
  \bibinfo{pages}{1935} (\bibinfo{year}{1995}).

\bibitem[{\citenamefont{Sastry et~al.}(1998)\citenamefont{Sastry, Debenedetti,
  and Stillinger}}]{Sastry98}
\bibinfo{author}{\bibfnamefont{S.}~\bibnamefont{Sastry}},
  \bibinfo{author}{\bibfnamefont{P.~G.} \bibnamefont{Debenedetti}},
  \bibnamefont{and} \bibinfo{author}{\bibfnamefont{R.~H.}
  \bibnamefont{Stillinger}}, \bibinfo{journal}{Nature}
  \textbf{\bibinfo{volume}{393}}, \bibinfo{pages}{554} (\bibinfo{year}{1998}).

\bibitem[{\citenamefont{Moynihan et~al.}(1976)\citenamefont{Moynihan, Easteal,
  DeBolt, and Tucker}}]{Moynihan76}
\bibinfo{author}{\bibfnamefont{C.~T.} \bibnamefont{Moynihan}},
  \bibinfo{author}{\bibfnamefont{A.~J.} \bibnamefont{Easteal}},
  \bibinfo{author}{\bibfnamefont{M.~A.} \bibnamefont{DeBolt}},
  \bibnamefont{and} \bibinfo{author}{\bibfnamefont{J.}~\bibnamefont{Tucker}},
  \bibinfo{journal}{J. Am. Ceram. Soc.} \textbf{\bibinfo{volume}{59}},
  \bibinfo{pages}{12} (\bibinfo{year}{1976}).

\bibitem[{\citenamefont{Adam and Gibbs}(1965)}]{Adam65}
\bibinfo{author}{\bibfnamefont{G.}~\bibnamefont{Adam}} \bibnamefont{and}
  \bibinfo{author}{\bibfnamefont{J.~H.} \bibnamefont{Gibbs}},
  \bibinfo{journal}{J. Chem. Phys.} \textbf{\bibinfo{volume}{43}},
  \bibinfo{pages}{139} (\bibinfo{year}{1965}).

\bibitem[{\citenamefont{Gupta and Mauro}(2007)}]{Gupta07}
\bibinfo{author}{\bibfnamefont{P.~K.} \bibnamefont{Gupta}} \bibnamefont{and}
  \bibinfo{author}{\bibfnamefont{J.~C.} \bibnamefont{Mauro}},
  \bibinfo{journal}{J. Chem. Phys.} \textbf{\bibinfo{volume}{126}},
  \bibinfo{pages}{224505} (\bibinfo{year}{2007}).

\bibitem[{\citenamefont{Hentschel et~al.}(2008)\citenamefont{Hentschel, Ilyin,
  Procassia, and Schupper}}]{Hentschel08}
\bibinfo{author}{\bibfnamefont{H.~G.~E.} \bibnamefont{Hentschel}},
  \bibinfo{author}{\bibfnamefont{V.}~\bibnamefont{Ilyin}},
  \bibinfo{author}{\bibfnamefont{I.}~\bibnamefont{Procassia}},
  \bibnamefont{and} \bibinfo{author}{\bibfnamefont{N.}~\bibnamefont{Schupper}},
  \bibinfo{journal}{Phys. Rev. E} \textbf{\bibinfo{volume}{78}},
  \bibinfo{pages}{061504} (\bibinfo{year}{2008}).

\bibitem[{\citenamefont{Oblad and Newton}(1937)}]{Oblad37}
\bibinfo{author}{\bibfnamefont{A.~G.} \bibnamefont{Oblad}} \bibnamefont{and}
  \bibinfo{author}{\bibfnamefont{R.~F.} \bibnamefont{Newton}},
  \bibinfo{journal}{J. Am. Chem. Soc.} \textbf{\bibinfo{volume}{59}},
  \bibinfo{pages}{2495} (\bibinfo{year}{1937}).

\bibitem[{\citenamefont{Porter et~al.}(2009)\citenamefont{Porter, Easterling,
  and Sherif}}]{Porter-PT-metals}
\bibinfo{author}{\bibfnamefont{D.~A.} \bibnamefont{Porter}},
  \bibinfo{author}{\bibfnamefont{K.~E.} \bibnamefont{Easterling}},
  \bibnamefont{and} \bibinfo{author}{\bibfnamefont{M.~Y.}
  \bibnamefont{Sherif}}, \emph{\bibinfo{title}{Phase Transformations in Metals
  and Alloys}} (\bibinfo{publisher}{CRC Press}, \bibinfo{address}{Boca Raton},
  \bibinfo{year}{2009}), \bibinfo{edition}{3rd} ed.

\bibitem[{\citenamefont{Bundy}(1985)}]{Bundy85}
\bibinfo{author}{\bibfnamefont{F.~P.} \bibnamefont{Bundy}}, in
  \emph{\bibinfo{booktitle}{Solid State Physics Under Pressure: Recent Advances
  with Anvil Devices}}, edited by
  \bibinfo{editor}{\bibfnamefont{S.}~\bibnamefont{Minomura}}
  (\bibinfo{publisher}{KTK Scientific}, \bibinfo{address}{Tokyo},
  \bibinfo{year}{1985}), p.~\bibinfo{pages}{1}.

\bibitem[{\citenamefont{Spitsyn et~al.}(1981)\citenamefont{Spitsyn, Bouilov,
  and Derjagoin}}]{Spitsyn81}
\bibinfo{author}{\bibfnamefont{B.~V.} \bibnamefont{Spitsyn}},
  \bibinfo{author}{\bibfnamefont{L.~L.} \bibnamefont{Bouilov}},
  \bibnamefont{and} \bibinfo{author}{\bibfnamefont{B.~V.}
  \bibnamefont{Derjagoin}}, \bibinfo{journal}{J. Cryst. Growth}
  \textbf{\bibinfo{volume}{52}}, \bibinfo{pages}{219} (\bibinfo{year}{1981}).

\bibitem[{\citenamefont{Kamo et~al.}(1983)\citenamefont{Kamo, Sato, Matsumoto,
  and Setaka}}]{Kamo83}
\bibinfo{author}{\bibfnamefont{M.}~\bibnamefont{Kamo}},
  \bibinfo{author}{\bibfnamefont{Y.}~\bibnamefont{Sato}},
  \bibinfo{author}{\bibfnamefont{S.}~\bibnamefont{Matsumoto}},
  \bibnamefont{and} \bibinfo{author}{\bibfnamefont{N.}~\bibnamefont{Setaka}},
  \bibinfo{journal}{J. Cryst. Growth} \textbf{\bibinfo{volume}{62}},
  \bibinfo{pages}{642} (\bibinfo{year}{1983}).

\bibitem[{\citenamefont{Uemura et~al.}(2019)\citenamefont{Uemura, Shirai,
  Kunstmann, Ekimov, and Lebed}}]{Uemura19}
\bibinfo{author}{\bibfnamefont{N.}~\bibnamefont{Uemura}},
  \bibinfo{author}{\bibfnamefont{K.}~\bibnamefont{Shirai}},
  \bibinfo{author}{\bibfnamefont{J.}~\bibnamefont{Kunstmann}},
  \bibinfo{author}{\bibfnamefont{E.~A.} \bibnamefont{Ekimov}},
  \bibnamefont{and} \bibinfo{author}{\bibfnamefont{Y.~B.} \bibnamefont{Lebed}},
  \bibinfo{journal}{J. Phys.: Mater.} \textbf{\bibinfo{volume}{2}},
  \bibinfo{pages}{045004} (\bibinfo{year}{2019}).

\bibitem[{\citenamefont{Galeener et~al.}(1983)\citenamefont{Galeener,
  Leadbetter, and Stringfellow}}]{Galeener83}
\bibinfo{author}{\bibfnamefont{F.~L.} \bibnamefont{Galeener}},
  \bibinfo{author}{\bibfnamefont{A.~J.} \bibnamefont{Leadbetter}},
  \bibnamefont{and} \bibinfo{author}{\bibfnamefont{M.~W.}
  \bibnamefont{Stringfellow}}, \bibinfo{journal}{Phys. Rev. B}
  \textbf{\bibinfo{volume}{27}}, \bibinfo{pages}{1052} (\bibinfo{year}{1983}).

\bibitem[{\citenamefont{Sen and Thorpe}(1977)}]{Sen77}
\bibinfo{author}{\bibfnamefont{P.~N.} \bibnamefont{Sen}} \bibnamefont{and}
  \bibinfo{author}{\bibfnamefont{M.~F.} \bibnamefont{Thorpe}},
  \bibinfo{journal}{Phys. Rev. B} \textbf{\bibinfo{volume}{15}},
  \bibinfo{pages}{4030} (\bibinfo{year}{1977}).

\bibitem[{\citenamefont{\v{S}tich et~al.}(1991)\citenamefont{\v{S}tich, Car,
  and Parrinello}}]{Stich91}
\bibinfo{author}{\bibfnamefont{I.}~\bibnamefont{\v{S}tich}},
  \bibinfo{author}{\bibfnamefont{R.}~\bibnamefont{Car}}, \bibnamefont{and}
  \bibinfo{author}{\bibfnamefont{M.}~\bibnamefont{Parrinello}},
  \bibinfo{journal}{Phys. Rev. B} \textbf{\bibinfo{volume}{44}},
  \bibinfo{pages}{11092} (\bibinfo{year}{1991}).

\bibitem[{\citenamefont{Wilson}(1957)}]{Wilson}
\bibinfo{author}{\bibfnamefont{A.~H.} \bibnamefont{Wilson}},
  \emph{\bibinfo{title}{Thermodynamics and Statistical Thermodynamics}}
  (\bibinfo{publisher}{Cambridge}, \bibinfo{address}{Cambridge},
  \bibinfo{year}{1957}).

\bibitem[{\citenamefont{Wilks}(1961)}]{Wilks}
\bibinfo{author}{\bibfnamefont{J.}~\bibnamefont{Wilks}},
  \emph{\bibinfo{title}{The Third Law of Thermodynamics}}
  (\bibinfo{publisher}{Oxford}, \bibinfo{address}{London},
  \bibinfo{year}{1961}).

\bibitem[{Shi()}]{Shirai18-res}
\bibinfo{note}{K. Shirai, cond-mat.stat-mech/1804.02122}.

\bibitem[{\citenamefont{Denbigh}(1989)}]{Denbigh89}
\bibinfo{author}{\bibfnamefont{K.~G.} \bibnamefont{Denbigh}},
  \bibinfo{journal}{Brit. J. Phil. Sci.} \textbf{\bibinfo{volume}{40}},
  \bibinfo{pages}{323} (\bibinfo{year}{1989}).

\bibitem[{\citenamefont{Styer}(2000)}]{Styer00}
\bibinfo{author}{\bibfnamefont{D.~F.} \bibnamefont{Styer}},
  \bibinfo{journal}{Am. J. Phys} \textbf{\bibinfo{volume}{68}},
  \bibinfo{pages}{1090} (\bibinfo{year}{2000}).

\bibitem[{\citenamefont{Gibson and Giauque}(1923)}]{Gibson23}
\bibinfo{author}{\bibfnamefont{G.~E.} \bibnamefont{Gibson}} \bibnamefont{and}
  \bibinfo{author}{\bibfnamefont{W.~F.} \bibnamefont{Giauque}},
  \bibinfo{journal}{J. Am. Chem. Soc.} \textbf{\bibinfo{volume}{45}},
  \bibinfo{pages}{93} (\bibinfo{year}{1923}).

\bibitem[{\citenamefont{Simon and Lange}(1926)}]{Simon26}
\bibinfo{author}{\bibfnamefont{F.}~\bibnamefont{Simon}} \bibnamefont{and}
  \bibinfo{author}{\bibfnamefont{F.}~\bibnamefont{Lange}},
  \bibinfo{journal}{Zeit. f. Phys.} \textbf{\bibinfo{volume}{38}},
  \bibinfo{pages}{227} (\bibinfo{year}{1926}).

\bibitem[{\citenamefont{Park et~al.}(1928)\citenamefont{Park, Fuffman, and
  Cattoir}}]{Park28}
\bibinfo{author}{\bibfnamefont{G.~S.} \bibnamefont{Park}},
  \bibinfo{author}{\bibfnamefont{H.~M.} \bibnamefont{Fuffman}},
  \bibnamefont{and} \bibinfo{author}{\bibfnamefont{F.~R.}
  \bibnamefont{Cattoir}}, \bibinfo{journal}{J. Phys. Chem.}
  \textbf{\bibinfo{volume}{32}}, \bibinfo{pages}{1366} (\bibinfo{year}{1928}).

\bibitem[{\citenamefont{Chang and Bestul}(1972)}]{Chang72}
\bibinfo{author}{\bibfnamefont{S.~S.} \bibnamefont{Chang}} \bibnamefont{and}
  \bibinfo{author}{\bibfnamefont{A.~B.} \bibnamefont{Bestul}},
  \bibinfo{journal}{J. Chem. Phys.} \textbf{\bibinfo{volume}{56}},
  \bibinfo{pages}{503} (\bibinfo{year}{1972}).

\bibitem[{\citenamefont{Gupta and Moynihan}(1976)}]{Gupta76}
\bibinfo{author}{\bibfnamefont{P.~K.} \bibnamefont{Gupta}} \bibnamefont{and}
  \bibinfo{author}{\bibfnamefont{C.~T.} \bibnamefont{Moynihan}},
  \bibinfo{journal}{J. Chem. Phys.} \textbf{\bibinfo{volume}{65}},
  \bibinfo{pages}{4136} (\bibinfo{year}{1976}).

\bibitem[{\citenamefont{Lesikar and Moynihan}(1980)}]{Lesikar80}
\bibinfo{author}{\bibfnamefont{A.~V.} \bibnamefont{Lesikar}} \bibnamefont{and}
  \bibinfo{author}{\bibfnamefont{C.~T.} \bibnamefont{Moynihan}},
  \bibinfo{journal}{J. Chem. Phys.} \textbf{\bibinfo{volume}{73}},
  \bibinfo{pages}{1932} (\bibinfo{year}{1980}).

\bibitem[{\citenamefont{Franz and Parisi}(1997)}]{Franz97}
\bibinfo{author}{\bibfnamefont{S.}~\bibnamefont{Franz}} \bibnamefont{and}
  \bibinfo{author}{\bibfnamefont{G.}~\bibnamefont{Parisi}},
  \bibinfo{journal}{Phys. Rev. Lett.} \textbf{\bibinfo{volume}{79}},
  \bibinfo{pages}{2486} (\bibinfo{year}{1997}).

\bibitem[{\citenamefont{Xia and Wolynes}(2000)}]{Xia00}
\bibinfo{author}{\bibfnamefont{X.}~\bibnamefont{Xia}} \bibnamefont{and}
  \bibinfo{author}{\bibfnamefont{P.~G.} \bibnamefont{Wolynes}},
  \bibinfo{journal}{Proc. Nat. Acad. Sci.} \textbf{\bibinfo{volume}{97}},
  \bibinfo{pages}{2990} (\bibinfo{year}{2000}).

\bibitem[{\citenamefont{Charbonneau et~al.}(2014)\citenamefont{Charbonneau,
  Kurchan, Parisi, Urbani, and Zamponi}}]{Charbonneau14}
\bibinfo{author}{\bibfnamefont{P.}~\bibnamefont{Charbonneau}},
  \bibinfo{author}{\bibfnamefont{J.}~\bibnamefont{Kurchan}},
  \bibinfo{author}{\bibfnamefont{G.}~\bibnamefont{Parisi}},
  \bibinfo{author}{\bibfnamefont{P.}~\bibnamefont{Urbani}}, \bibnamefont{and}
  \bibinfo{author}{\bibfnamefont{F.}~\bibnamefont{Zamponi}},
  \bibinfo{journal}{Nat. Commun.} \textbf{\bibinfo{volume}{5:3725}},
  \bibinfo{pages}{1} (\bibinfo{year}{2014}).

\bibitem[{\citenamefont{Ritland}(1954)}]{Ritland54}
\bibinfo{author}{\bibfnamefont{H.~N.} \bibnamefont{Ritland}},
  \bibinfo{journal}{J. Am. Ceram. Soc.} \textbf{\bibinfo{volume}{37}},
  \bibinfo{pages}{370} (\bibinfo{year}{1954}).

\bibitem[{\citenamefont{Narayanaswamy}(1971)}]{Narayanaswamy71}
\bibinfo{author}{\bibfnamefont{O.~S.} \bibnamefont{Narayanaswamy}},
  \bibinfo{journal}{J. Am. Ceram. Soc.} \textbf{\bibinfo{volume}{54}},
  \bibinfo{pages}{491} (\bibinfo{year}{1971}).

\bibitem[{\citenamefont{Moynihan et~al.}(1974)\citenamefont{Moynihan, Easteal,
  Wilder, and Tucker}}]{Moynihan74}
\bibinfo{author}{\bibfnamefont{C.~T.} \bibnamefont{Moynihan}},
  \bibinfo{author}{\bibfnamefont{A.~J.} \bibnamefont{Easteal}},
  \bibinfo{author}{\bibfnamefont{J.}~\bibnamefont{Wilder}}, \bibnamefont{and}
  \bibinfo{author}{\bibfnamefont{J.}~\bibnamefont{Tucker}},
  \bibinfo{journal}{J. Phys. Chem.} \textbf{\bibinfo{volume}{78}},
  \bibinfo{pages}{2673} (\bibinfo{year}{1974}).

\bibitem[{\citenamefont{Kelton}(1991)}]{Kelton91}
\bibinfo{author}{\bibfnamefont{K.~F.} \bibnamefont{Kelton}},
  \bibinfo{journal}{Solid State Phys.} \textbf{\bibinfo{volume}{45}},
  \bibinfo{pages}{75} (\bibinfo{year}{1991}).

\bibitem[{\citenamefont{Debolt et~al.}(1976)\citenamefont{Debolt, Easteal,
  Macedo, and Moynihan}}]{Debolt76}
\bibinfo{author}{\bibfnamefont{M.~A.} \bibnamefont{Debolt}},
  \bibinfo{author}{\bibfnamefont{A.~J.} \bibnamefont{Easteal}},
  \bibinfo{author}{\bibfnamefont{P.~B.} \bibnamefont{Macedo}},
  \bibnamefont{and} \bibinfo{author}{\bibfnamefont{C.~T.}
  \bibnamefont{Moynihan}}, \bibinfo{journal}{J. Am. Ceram. Soc.}
  \textbf{\bibinfo{volume}{59}}, \bibinfo{pages}{16} (\bibinfo{year}{1976}).

\bibitem[{\citenamefont{Hodge}(1983)}]{Hodge83}
\bibinfo{author}{\bibfnamefont{I.~M.} \bibnamefont{Hodge}},
  \bibinfo{journal}{Macromolecules} \textbf{\bibinfo{volume}{16}},
  \bibinfo{pages}{898} (\bibinfo{year}{1983}).

\bibitem[{\citenamefont{Hutchinson and Kovacs}(1976)}]{Hutchinson76}
\bibinfo{author}{\bibfnamefont{J.~M.} \bibnamefont{Hutchinson}}
  \bibnamefont{and} \bibinfo{author}{\bibfnamefont{A.~J.}
  \bibnamefont{Kovacs}}, \bibinfo{journal}{J. Polym. Sci.: Polym. Phys. Ed.}
  \textbf{\bibinfo{volume}{14}}, \bibinfo{pages}{1575} (\bibinfo{year}{1976}).

\bibitem[{\citenamefont{Kovacs et~al.}(1979)\citenamefont{Kovacs, Aklonis,
  Hutchinson, and Ramos}}]{Kovacs79}
\bibinfo{author}{\bibfnamefont{A.~J.} \bibnamefont{Kovacs}},
  \bibinfo{author}{\bibfnamefont{J.~J.} \bibnamefont{Aklonis}},
  \bibinfo{author}{\bibfnamefont{J.~M.} \bibnamefont{Hutchinson}},
  \bibnamefont{and} \bibinfo{author}{\bibfnamefont{A.~R.} \bibnamefont{Ramos}},
  \bibinfo{journal}{J. Polym. Sci.: Polym. Phys. Ed.}
  \textbf{\bibinfo{volume}{17}}, \bibinfo{pages}{1097} (\bibinfo{year}{1979}).

\bibitem[{\citenamefont{Lillie}(1936)}]{Lillie36}
\bibinfo{author}{\bibfnamefont{H.~R.} \bibnamefont{Lillie}},
  \bibinfo{journal}{J. Am. Ceram. Soc.} \textbf{\bibinfo{volume}{19}},
  \bibinfo{pages}{45} (\bibinfo{year}{1936}).

\bibitem[{\citenamefont{Ritland}(1956)}]{Ritland56}
\bibinfo{author}{\bibfnamefont{H.~N.} \bibnamefont{Ritland}},
  \bibinfo{journal}{J. Am. Ceram. Soc.} \textbf{\bibinfo{volume}{39}},
  \bibinfo{pages}{403} (\bibinfo{year}{1956}).

\bibitem[{\citenamefont{Gardon and Narayanaswamy}(1970)}]{Gardon70}
\bibinfo{author}{\bibfnamefont{R.}~\bibnamefont{Gardon}} \bibnamefont{and}
  \bibinfo{author}{\bibfnamefont{O.~S.} \bibnamefont{Narayanaswamy}},
  \bibinfo{journal}{J. Am. Ceram. Soc.} \textbf{\bibinfo{volume}{53}},
  \bibinfo{pages}{380} (\bibinfo{year}{1970}).

\bibitem[{\citenamefont{Tool}(1946)}]{Tool46}
\bibinfo{author}{\bibfnamefont{A.~Q.} \bibnamefont{Tool}}, \bibinfo{journal}{J.
  Am. Ceram. Soc.} \textbf{\bibinfo{volume}{29}}, \bibinfo{pages}{240}
  (\bibinfo{year}{1946}).

\bibitem[{\citenamefont{Moynihan et~al.}(1981)\citenamefont{Moynihan, Macedo,
  Montrose, Gupta, Debolt, Dill, Dom, Drake, Easteal, Elterman
  et~al.}}]{Moynihan76a}
\bibinfo{author}{\bibfnamefont{C.~T.} \bibnamefont{Moynihan}},
  \bibinfo{author}{\bibfnamefont{P.~B.} \bibnamefont{Macedo}},
  \bibinfo{author}{\bibfnamefont{C.~H.} \bibnamefont{Montrose}},
  \bibinfo{author}{\bibfnamefont{P.~K.} \bibnamefont{Gupta}},
  \bibinfo{author}{\bibfnamefont{M.~A.} \bibnamefont{Debolt}},
  \bibinfo{author}{\bibfnamefont{J.~F.} \bibnamefont{Dill}},
  \bibinfo{author}{\bibfnamefont{B.~E.} \bibnamefont{Dom}},
  \bibinfo{author}{\bibfnamefont{P.~Z.} \bibnamefont{Drake}},
  \bibinfo{author}{\bibfnamefont{A.~J.} \bibnamefont{Easteal}},
  \bibinfo{author}{\bibfnamefont{P.~B.} \bibnamefont{Elterman}},
  \bibnamefont{et~al.}, \bibinfo{journal}{Ann.~New York Acad.~Sci.}
  \textbf{\bibinfo{volume}{371}}, \bibinfo{pages}{151} (\bibinfo{year}{1981}).

\bibitem[{\citenamefont{Hodge}(1994)}]{Hodge94}
\bibinfo{author}{\bibfnamefont{I.~M.} \bibnamefont{Hodge}},
  \bibinfo{journal}{J. Non-Cryst. Solids} \textbf{\bibinfo{volume}{169}},
  \bibinfo{pages}{211} (\bibinfo{year}{1994}).

\bibitem[{\citenamefont{Debenedetti and Stillinger}(2001)}]{Debenedetti01}
\bibinfo{author}{\bibfnamefont{P.~G.} \bibnamefont{Debenedetti}}
  \bibnamefont{and} \bibinfo{author}{\bibfnamefont{F.~H.}
  \bibnamefont{Stillinger}}, \bibinfo{journal}{Nature}
  \textbf{\bibinfo{volume}{410}}, \bibinfo{pages}{259} (\bibinfo{year}{2001}).

\bibitem[{\citenamefont{Angell}(1988)}]{Angell88}
\bibinfo{author}{\bibfnamefont{C.~A.} \bibnamefont{Angell}},
  \bibinfo{journal}{J. Phys. Chem. Solids} \textbf{\bibinfo{volume}{49}},
  \bibinfo{pages}{863} (\bibinfo{year}{1988}).

\bibitem[{\citenamefont{Angell}(1995)}]{Angell95}
\bibinfo{author}{\bibfnamefont{C.~A.} \bibnamefont{Angell}},
  \bibinfo{journal}{Science} \textbf{\bibinfo{volume}{267}},
  \bibinfo{pages}{1924} (\bibinfo{year}{1995}).

\bibitem[{\citenamefont{Scherer}(1984)}]{Scherer84}
\bibinfo{author}{\bibfnamefont{G.~W.} \bibnamefont{Scherer}},
  \bibinfo{journal}{J. Am. Ceram. Soc.} \textbf{\bibinfo{volume}{67}},
  \bibinfo{pages}{504} (\bibinfo{year}{1984}).

\bibitem[{\citenamefont{Nemilov}(1995)}]{Nemilov-VitreousState}
\bibinfo{author}{\bibfnamefont{S.~V.} \bibnamefont{Nemilov}},
  \emph{\bibinfo{title}{Thermodynamic and Kinetic Aspects of the Vitreous
  State}} (\bibinfo{publisher}{CRC Press}, \bibinfo{address}{Boca Raton},
  \bibinfo{year}{1995}).

\bibitem[{\citenamefont{Johari}(2000)}]{Johari00}
\bibinfo{author}{\bibfnamefont{G.~P.} \bibnamefont{Johari}},
  \bibinfo{journal}{J. Chem. Phys.} \textbf{\bibinfo{volume}{113}},
  \bibinfo{pages}{8958} (\bibinfo{year}{2000}).

\bibitem[{\citenamefont{Sastry}(2001)}]{Sastry01}
\bibinfo{author}{\bibfnamefont{S.}~\bibnamefont{Sastry}},
  \bibinfo{journal}{Nature} \textbf{\bibinfo{volume}{409}},
  \bibinfo{pages}{164} (\bibinfo{year}{2001}).

\bibitem[{\citenamefont{Svoboda and M\'{a}lek}(2013)}]{Svoboda13}
\bibinfo{author}{\bibfnamefont{R.}~\bibnamefont{Svoboda}} \bibnamefont{and}
  \bibinfo{author}{\bibfnamefont{J.}~\bibnamefont{M\'{a}lek}},
  \bibinfo{journal}{Polymer} \textbf{\bibinfo{volume}{54}},
  \bibinfo{pages}{1504} (\bibinfo{year}{2013}).

\bibitem[{\citenamefont{Hodge}(1991)}]{Hodge91}
\bibinfo{author}{\bibfnamefont{I.~M.} \bibnamefont{Hodge}},
  \bibinfo{journal}{J. Non-Cryst. Solids} \textbf{\bibinfo{volume}{131-133}},
  \bibinfo{pages}{435} (\bibinfo{year}{1991}).

\bibitem[{\citenamefont{Mehrer}(2007)}]{Mehrer07}
\bibinfo{author}{\bibfnamefont{H.}~\bibnamefont{Mehrer}},
  \emph{\bibinfo{title}{Diffusion in Solids: Fundamentals, Methods, Materials,
  Diffusion-Controlled Processes}} (\bibinfo{publisher}{Springer},
  \bibinfo{address}{Berlin}, \bibinfo{year}{2007}).

\bibitem[{\citenamefont{Fahey et~al.}(1989)\citenamefont{Fahey, Griffin, and
  Plummer}}]{Fahey89}
\bibinfo{author}{\bibfnamefont{P.~M.} \bibnamefont{Fahey}},
  \bibinfo{author}{\bibfnamefont{P.~B.} \bibnamefont{Griffin}},
  \bibnamefont{and} \bibinfo{author}{\bibfnamefont{J.~D.}
  \bibnamefont{Plummer}}, \bibinfo{journal}{Rev. Mod. Phys.}
  \textbf{\bibinfo{volume}{61}}, \bibinfo{pages}{289} (\bibinfo{year}{1989}).

\bibitem[{\citenamefont{Parks et~al.}(1928)\citenamefont{Parks, Huffman, and
  Cattoir}}]{Parks28}
\bibinfo{author}{\bibfnamefont{G.}~\bibnamefont{Parks}},
  \bibinfo{author}{\bibnamefont{Huffman}}, \bibnamefont{and}
  \bibinfo{author}{\bibnamefont{Cattoir}}, \bibinfo{journal}{J. Phys. Chem.}
  \textbf{\bibinfo{volume}{32}}, \bibinfo{pages}{1366} (\bibinfo{year}{1928}).

\bibitem[{\citenamefont{Parks and Thomas}(1934)}]{Parks34}
\bibinfo{author}{\bibfnamefont{G.}~\bibnamefont{Parks}} \bibnamefont{and}
  \bibinfo{author}{\bibfnamefont{S.~B.} \bibnamefont{Thomas}},
  \bibinfo{journal}{J. Am. Chem. Soc.} \textbf{\bibinfo{volume}{56}},
  \bibinfo{pages}{1423} (\bibinfo{year}{1934}).

\end{thebibliography}

\end{document}